\documentclass[useAMS,usenatbib]{mn2e}
\usepackage{amssymb,amsmath,graphicx,color}

\usepackage{url}
\usepackage{array}
\title[Photospheric emission from collimated outflows]{A theory of photospheric emission from relativistic, collimated outflows}
\author[C. Lundman, A. Pe'er and F. Ryde]{C. Lundman$^{1,2}$\thanks{E-mail:
clundman@particle.kth.se (CL); apeer@cfa.harvard.edu (AP); fryde@particle.kth.se (FR)}, A. Pe'er$^{3}$\footnotemark[1] and F. Ryde$^{1,2}$\footnotemark[1] \\
$^{1}$Department of Physics, Royal Institute of Technology (KTH), AlbaNova, SE-106 91 Stockholm, Sweden \\
$^{2}$The Oskar Klein Centre for Cosmoparticle Physics, AlbaNova, SE-106 91 Stockholm, Sweden \\
$^{3}$Harvard-Smithsonian Center for Astrophysics, Cambridge, MA 02138, USA}
\begin{document}

\date{Accepted --}

\pagerange{\pageref{firstpage}--\pageref{lastpage}} \pubyear{2012}

\maketitle

\label{firstpage}

\begin{abstract}
Relativistic outflows in the form of jets are common in many astrophysical objects. By their very nature, jets have angle dependent velocity profiles, $\Gamma = \Gamma(r,\theta,\phi)$, where $\Gamma$ is the outflow Lorentz factor. In this work we consider photospheric emission from non-dissipative jets with various Lorentz factor profiles, of the approximate form $\Gamma \approx \Gamma_{\rm 0}/[(\theta/\theta_{\rm j})^p + 1]$, where $\theta_{\rm j}$ is the characteristic jet opening angle. In collimated jets, the observed spectrum depends on the viewing angle, $\theta_{\rm v}$. We show that for narrow jets ($\theta_{\rm j} \Gamma_{\rm 0} \lesssim few$), the obtained low energy photon index is $\alpha \approx -1$ ($dN/dE \propto E^\alpha$), independent of viewing angle, and weakly dependent on the Lorentz factor gradient ($p$). A similar result is obtained for wider jets observed at $\theta_{\rm v} \approx \theta_{\rm j}$. This result is surprisingly similar to the average low energy photon index seen in gamma-ray bursts. For wide jets ($\theta_{\rm j} \Gamma_{\rm 0} \gtrsim few$) observed at $\theta_{\rm v} \ll \theta_{\rm j}$, a multicolor blackbody spectrum is obtained. We discuss the consequences of this theory on our understanding of the prompt emission in gamma-ray bursts.
\end{abstract}


\section{Introduction}
\label{sect:introduction}

Photospheric emission from highly relativistic outflows was early considered as an explanation for prompt gamma-ray bursts (GRBs, \citealt{Goo:1986, Pac:1986}). It is a natural consequence of the fireball model, where the optical depth at the base of the outflow is much larger than unity (e.g. \citealt{Pir:1999}). Moreover, photospheric emission provides a natural explanation to the small dispersion of the sub-MeV peak and to the high prompt emission efficiency observed \citep{MesRee:2000}. However, the observed spectrum usually appears significantly broader than a Planck spectrum \citep{PreEtAl:2000}, being well fitted by a smoothly broken power law (the Band function, \citealt{BanEtAl:1993}). Thus, the prompt emission has commonly been associated with synchrotron emission originating from kinetic energy dissipation outside the photosphere \citep{ReeMes:1994}. However, in recent years it has become clear that optically thin synchrotron emission is incompatible with the hard low energy slopes observed in a substantial fraction of GRBs \citep{PreEtAl:1998, KanEtAl:2006, Bel:2012}. This has raised the need for alternative ideas.

One appealing idea is broadening of the thermal spectrum emitted from the photosphere. The emerging photon spectrum from a static, optically thick, relativistic outflow can be widened in two ways:

\begin{enumerate}
\item Energy dissipation below the photosphere can heat electrons above the equillibrium temperature. These electrons emit synchrotron emission and Comptonize the thermal photons, thereby modifying the Planck spectrum \citep{ReeMes:2005, PeeMesRee:2005, PeeMesRee:2006}. The dissipation can be caused by shocks \citep{ReeMes:2005, LazMorBeg:2009, RydEtAl:2011}, dissipation of magnetic energy \citep{Tho:1994, SprDaiDre:2001, GiaSpr:2005, ZhaYan:2011} or collisional processes \citep{Bel:2010, VurEtAl:2011}.
\item The photospheric radius is angle dependent \citep{AbrNovPac:1991, Pee:2008}. Moreover, it was shown by \citet{Pee:2008} that photons make their last scatterings at a distribution of radii and angles. The observer sees simultaneously photons emitted from a large range of radii and angles. Therefore, the observed spectrum is a superposition of comoving spectra. The Doppler boost is a function of angle, and the comoving temperature decreases with radius through adiabatic cooling. Depending on the outflow properties, this geometrical broadening can form observed spectra which appears significantly different from the Planck spectrum.
\end{enumerate}

Photospheric emission in the context of spherically symmetric outflows has been studied by several authors. \citet{Goo:1986} considered a highly relativistic outflow in the context of cosmological GRBs. It was realized that the observed spectrum is broader than blackbody, however the analysis was one-dimensional. \citet{AbrNovPac:1991} realized that the two-dimensional shape of the photosphere in a relativistic, spherically symmetric wind is in fact concave and symmetric around the line-of-sight (LOS). This can be understood as a consequence of the dependence on viewing angle of the optical depth of relativistically moving media. \citet{Pee:2008} found a simple expression for the photospheric radius, $R_{\rm ph} \propto (\theta^2/3 + 1/\Gamma^2)$, where $\theta$ is the angle measured from the LOS, $\Gamma \equiv (1 - \beta)^{-1/2}$ is the outflow bulk Lorentz factor and $\beta$ is the outflow speed in units of the speed of light. \citet{Pee:2008} extended the photospheric emission model by recognizing the importance of considering photons from the entire emitting volume, introducing probability density distributions for the last scattering photon positions. \citet{Bel:2011} took a different approach, solving the radiative transfer equation in the relativistic limit. The approximate probability densities used by \citet{Pee:2008} was validated by \citet{Bel:2011}.

All above mentioned works considered spherical explosions. As we show below, this is a good approximation for collimated outflows as long as the characteristic jet opening angle is much larger than $1/\Gamma$ and the outflow is observed at viewing angles much smaller than the jet opening angle. Within the collapsar model \citep{MacWoo:1999} the jet is collimated by the pressure of the surrounding gas as it drills its way through the collapsing progenitor star. In such jets the position of the observer relative to the jet axis can affect the observed spectrum. Here we develop the theory of photospheric emission in collimated outflows, calculating the expression for the observed spectrum at any viewing angle.

The mechanism responsible for jet collimation is not fully understood. Hydrodynamic simulations of GRB jets after the launching phase (e.g., \citealt{ZhaWooMac:2003, MorLazBeg:2007, MizNagAoi:2011}) show the jet drilling through the stellar envelope, pushing material towards the sides, forming a hot cocoon (e.g. \citealt{AloEtAl:2000}). The surrounding gas then acts to collimate the jet. \citet{ZhaWooMac:2003} extracted angular profiles of the local rest mass density, total energy flux and outflow Lorentz factor at certain radii from the simulations. The resulting profiles can in many cases be approximated as constant within a characteristic jet opening angle, then decreasing as power laws towards the jet edge. In this work we adopt a similar parametrization of the angular profile of the bulk jet Lorentz factor, with a characteristic jet opening angle, power law index and normalization as free profile parameters.

We develop a model for photon propagation in the context of a steady, optically thick, axisymmetric jet with angle dependent electron number density, photon number density and bulk Lorentz factor. We compute the observed spectrum taking into account contributions from the entire emitting volume as seen by an observer located at any viewing angle. As an example solution, we consider fireball dynamics as a way to relate the angle dependent parameters, in combination with the assumed angular Lorentz factor profile. We developed a Monte Carlo simulation, unique in its ability to calculate photon propagation in a non-spherical explosion. We use this simulation to analyze the importance of angular bulk photon diffusion.

We show that for a large region in the parameter space, the low energy photon index is approximately $\alpha \approx -1$ (where $dN/dE \propto E^\alpha$). This is similar to the average value observed in GRBs \citep{KanEtAl:2006, NavEtAl:2011, GolEtAl:2012}. Furthermore, we present analytical expressions for the important energies and photon indices of the observed spectrum as functions of the free model parameters. We show that photospheric emission by itself can account for observations of GRB spectra below the peak energy without the need for energy dissipation below the photosphere or additional radiative processes, provided that the characteristic jet opening angle is not much larger than $\sim few/\Gamma$ or if the outflow is viewed off-axis. Although we focus on GRB parameters, the results are general and can readily be applied to any optically thick, relativistic outflow with lateral outflow properties such as active galactic nuclei.

This paper is organized as follows. In \S \ref{sect:model} we develop the analytical expression for the observed spectrum. We qualitatively explain the spectral features in \S \ref{sect:the observed spectrum} in terms of contributions from different jet regions. The Monte Carlo code is explained in \S \ref{sect:numerical simulations}, and in \S \ref{sect:results} we present both simulated and numerically integrated spectra for outflow parameters relevant to GRBs. We discuss model sensitivity and photon time delays in \S \ref{sect:discussion} and summarize our results in \S \ref{sect:summary and conclusions}.

%%%%%%%%%%%%%%%%%%%%%%%%%%%%%%%%%%%%%%%%%%%%%%%%
%%%%%%%%%%%%%%%%%%%%%%%%%%%%%%%%%%%%%%%%%%%%%%%%
%%%%%%%%%%%%%%%%%%%%%%%%%%%%%%%%%%%%%%%%%%%%%%%%
%%%%%%%%%%%%%%%%%%%%%%%%%%%%%%%%%%%%%%%%%%%%%%%%
%%%%%%%%%%%%%%%%%%%%%%%%%%%%%%%%%%%%%%%%%%%%%%%%
%%%%%%%%%%%%%%%%%%%%%%%%%%%%%%%%%%%%%%%%%%%%%%%%
%%%%%%%%%%%%%%%%%%%%%%%%%%%%%%%%%%%%%%%%%%%%%%%%
%%%%%%%%%%%%%%%%%%%%%%%%%%%%%%%%%%%%%%%%%%%%%%%%
%%%%%%%%%%%%%%%%%%%%%%%%%%%%%%%%%%%%%%%%%%%%%%%%
%%%%%%%%%%%%%%%%%%%%%%%%%%%%%%%%%%%%%%%%%%%%%%%%

\section{Model: Photon scattering in a steady jet}
\label{sect:model}

Motivated by the results of \citet{ZhaWooMac:2003}, we choose to define the shape of the angular Lorentz factor profile with an equation of the form

\begin{equation}
\left(\Gamma - \Gamma_{\rm min}\right)^2 = \frac{\left(\Gamma_{\rm 0} - \Gamma_{\rm min}\right)^2}{\left(\theta/\theta_{\rm j}\right)^{2p} + 1},
\label{eq:gamma profile}
\end{equation}

\begin{figure}
\includegraphics[width=\linewidth]{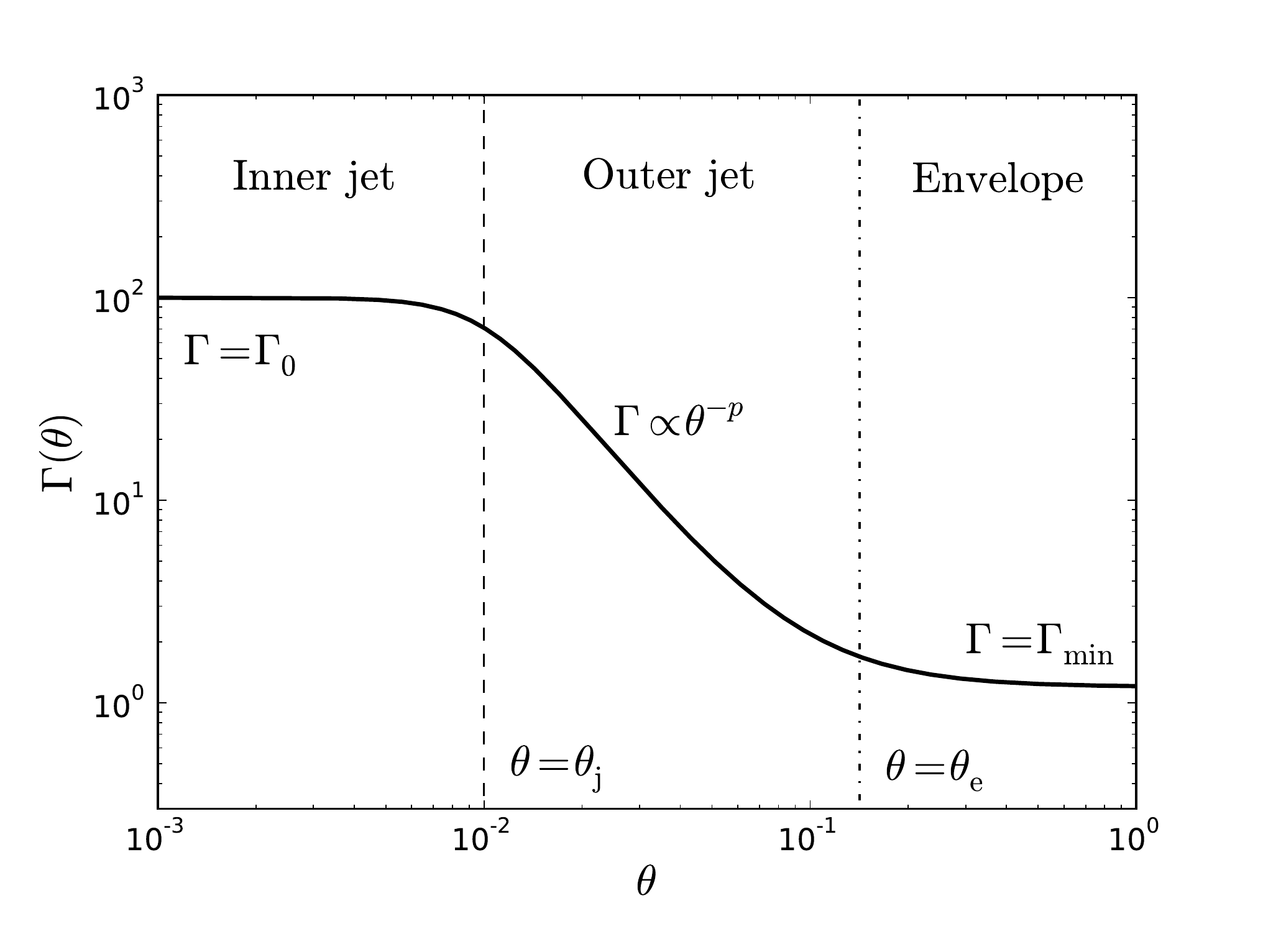}
\caption{An example Lorentz factor profile (Eq. \ref{eq:gamma profile}). In the inner jet ($\theta < \theta_{\rm j}$) the Lorentz factor is approximately constant, $\Gamma = \Gamma_{\rm 0}$ while in the outer jet ($\theta_{\rm j} < \theta < \theta_{\rm e}$) the Lorentz factor decrease is approximately that of a power law with index $-p$. In the envelope ($\theta > \theta_{\rm e}$), the Lorentz factor is approximately constant with a value of $\Gamma_{\rm min} = 1.2$. The dashed vertical line indicates $\theta_{\rm j}$, while the dash-dotted line shows the location of $\theta_{\rm e} \approx \theta_{\rm j} \Gamma_{\rm 0}^{1/p}$. For this figure $\Gamma_{\rm 0} = 100$, $\theta_{\rm j} = 0.01$ and $p = 2$.}
\label{fig:example lorentz factor}
\end{figure}

\noindent where $\Gamma_{\rm 0}$ is the maximum value of the Lorentz factor, $\theta_{\rm j}$ is the characteristic jet opening angle, $p$ determines the gradient of the profile and $\Gamma_{\rm min} = 1.2$ is the lowest value of the Lorentz factor, differing from unity for numerical reasons.\footnote{As shown in \S \ref{subsect:envelope component}, the exact value of $\Gamma_{\rm min}$ only affects the very low energy spectrum, many orders of magnitude below the observed peak energy.}

Three angular regions can be identified; the inner jet, the outer jet and the envelope. While this separation is artificial and a consequence of our chosen Lorentz factor profile, it is useful for understanding the resulting observed spectra. The inner jet is characterized by an approximately constant Lorentz factor, $\Gamma = \Gamma_{\rm 0}$. In the outer jet the Lorentz factor decreases approximately as a power law of the angle with index $p$, and in the envelope the Lorentz factor has an approximately constant value of $\Gamma \approx 1$. A characteristic angle separating the inner and outer jet regions is $\theta_{\rm j}$, where $\Gamma(\theta_{\rm j}) \approx \Gamma_{\rm 0}/\sqrt{2}$. Similarly, $\theta_{\rm e} \approx \theta_{\rm j} \Gamma_{\rm 0}^{1/p}$ is a characteristic angle separating the outer jet and the envelope regions.\footnote{In the calculations below we define $\theta_{\rm e}$ to be the angle where $\Gamma (\theta_{\rm e}) = \sqrt{2}\Gamma_{\rm min}$, which more accurately describes the characteristic angle separating the outer jet and the envelope regions. This gives $\theta_{\rm e} = \theta_{\rm j} [(\sqrt{2} + 1)\Gamma_{\rm 0}/\Gamma_{\rm min}]^{1/p}$.} An example Lorentz factor profile is shown in Figure \ref{fig:example lorentz factor}.

\begin{figure}
\includegraphics[width=\linewidth]{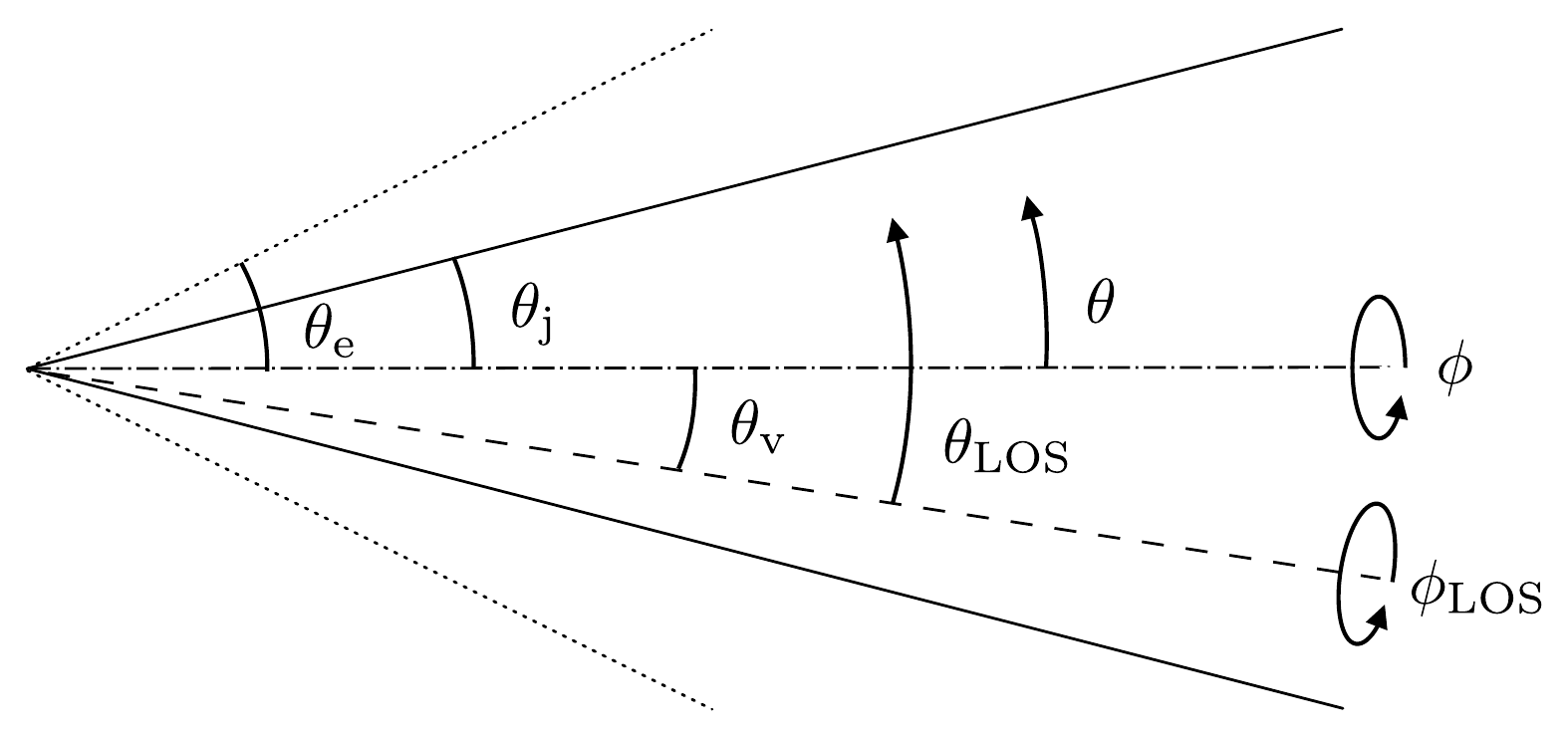}
\caption{Schematic view of the jet. The solid lines represent the characteristic jet opening angle, $\theta_{\rm j}$, as measured from the jet axis of symmetry, indicated by the dash-dotted line. The dotted line shows $\theta_{\rm e}$, the characteristic angle separating the outer jet and the envelope. The dashed line is the LOS, at an angle $\theta_{\rm v}$ to the jet axis. Note that in general, $\theta_{\rm LOS} \neq \theta + \theta_{\rm v}$.}
\label{fig:angle definitions}
\end{figure}

Physical properties of the jet are expressed in spherical coordinates $(r, \theta, \phi)$, with the polar axis aligned to the jet axis of symmetry (see Figure \ref{fig:angle definitions}). Some properties, such as the Doppler boost, depend on the angle to the LOS. These are most easily expressed in spherical coordinates $(r, \theta_{\rm LOS}, \phi_{\rm LOS})$, with the polar axis aligned to the LOS. {The radial coordinate ($r$) is measured along an axis that makes an angle $\theta$ with the jet axis and $\theta_{\rm LOS}$ with the LOS.} When viewing the jet head on, the two sets of coordinates coincide, and the radiative contributions from the jet are independent of azimuthal angle due to symmetry around the LOS. Our jet profile contains four parameters (including the viewing angle), three more than the simplified scenario of spherical symmetry.

%%%%%%%%%%%%%%%%%%%%%%%%%%%%%%%%%%%%%%%%%%%%%%%%
%%%%%%%%%%%%%%%%%%%%%%%%%%%%%%%%%%%%%%%%%%%%%%%%
%%%%%%%%%%%%%%%%%%%%%%%%%%%%%%%%%%%%%%%%%%%%%%%%
%%%%%%%%%%%%%%%%%%%%%%%%%%%%%%%%%%%%%%%%%%%%%%%%
%%%%%%%%%%%%%%%%%%%%%%%%%%%%%%%%%%%%%%%%%%%%%%%%

\subsection{The photosphere and the decoupling radius}
\label{subsect:the decoupling radius and the photosphere}

The photospheric radius, $R_{\rm ph}$, is defined as the radius which fulfills the following condition: The optical depth for a photon that propagates from that radius towards the observer is equal to unity. { Thus the photospheric radius defines a surface in space. The radial coordinate of this surface is $R_{\rm ph} = R_{\rm ph}(\theta_{\rm LOS}, \phi_{\rm LOS})$. From any point on this surface, the optical depth as measured towards the observer equals unity. In order for a photon to reach the observer, it has to propagate along a ray parallel to the LOS. Along such a ray, $r \sin\theta_{\rm LOS} = const$. We define the $z$-axis along this ray, $z \equiv r \cos\theta_{\rm LOS}$ is the spatial photon coordinate.} For the parameter values considered in this work, the comoving temperature at all relevant radii is much lower than the electron rest mass energy, and so we use the Thomson cross section in describing the scattering process. The optical depth from $z = z_{\rm min}$ to $z \rightarrow \infty$ is $\tau_{\rm ray} \equiv \int_{z_{\rm min}}^\infty n_{\rm e}^\prime \sigma_{\rm T} \mathcal{D}^{-1} dz$, where $\mathcal{D} = [\Gamma(1-\beta\cos\theta_{\rm LOS})]^{-1}$ is the Doppler boost, $n_{\rm e}^\prime$ is the comoving electron number density and $\sigma_{\rm T}$ is the Thomson cross section \citep{AbrNovPac:1991, Pee:2008}. The photospheric radius is a function of angle to the LOS, as follows from the definition of $\tau_{\rm ray}$. For a spherically symmetric outflow, due to symmetry, the shape of the photosphere is independent of viewing angle. In this work we consider jetted outflows, for which this is no longer true.

Here, in addition, we define the decoupling radius, $R_{\rm dcp}$, as the radius from which the optical depth for a photon that propagates in the {\it radial} direction is equal to unity. The optical depth from $r = r_{\rm min}$ to $r \rightarrow \infty$ is $\tau_{\rm rdl} \equiv \int_{r_{\rm min}}^\infty \Gamma (1 - \beta) n_{\rm e}^\prime \sigma_{\rm T} dr$. The importance of $R_{\rm dcp}$ is that this is the characteristic radius where photons and electrons fall out of thermal equillibrium. If the outflow has axisymmetric comoving electron density or bulk Lorentz factor, then $R_{\rm dcp}$ is a function of angle to the axis of symmetry. For photons propagating along the LOS ($\cos\theta_{\rm LOS} = 1$), $dz = dr$ and $R_{\rm ph} = R_{\rm dcp}$.

%%%%%%%%%%%%%%%%%%%%%%%%%%%%%%%%%%%%%%%%%%%%%%%%
%%%%%%%%%%%%%%%%%%%%%%%%%%%%%%%%%%%%%%%%%%%%%%%%
%%%%%%%%%%%%%%%%%%%%%%%%%%%%%%%%%%%%%%%%%%%%%%%%
%%%%%%%%%%%%%%%%%%%%%%%%%%%%%%%%%%%%%%%%%%%%%%%%
%%%%%%%%%%%%%%%%%%%%%%%%%%%%%%%%%%%%%%%%%%%%%%%%

\subsection{Formation of the observed spectrum}
\label{subsect:formation of the observed spectrum}

In order to calculate the observed spectrum we consider the following: while the photosphere is defined as $R_{\rm ph} = r(\tau_{\rm ray} = 1)$, in reality photons can scatter anywhere where electrons exist. Since we consider an expanding plasma in a steady state, electrons are assumed to occupy the entire volume.\footnote{Since the probability of scattering at $r \gg R_{\rm ph}$ is negligible, this is a good assumption at late times.} In order to calculate the observed spectrum one has to integrate the emissivity over the entire volume, where the emissivity in volume element $dV$ is proportional to the probability of photons to make their last scattering in $dV$.

We assume that the local, comoving photon energy distribution is well described by a Planck spectrum with temperature equal to the local, comoving electron temperature.\footnote{While this is true only for $r < R_{\rm dcp}$, as shown in \S \ref{subsect:outer jet component} this is a good assumption for most jet profiles considered in this work. We do not use this assumption in \S \ref{sect:numerical simulations}.} As we consider the photon-electron cross section to be energy independent, it follows that the optical depth is also energy independent. In this section we calculate the observed spectrum assuming that it depends only on the last scattering positions of the observed photons. We justify this assumption by simulations (\S \ref{sect:numerical simulations}) which consider full treatment of photon propagation below the photosphere.

Each time a photon in the outflow scatters into a direction pointing towards the observer, it has the probability $\exp(-\tau_{\rm ray})$ to escape the outflow. The rate of photons escaping the outflow from volume $dV$ into solid angle $d\Omega_{\rm v}$ is

\begin{equation}
d\dot{N}_{\rm esc} = \frac{d\dot{n}_{\rm sc}}{d\Omega_{\rm v}} \exp\left(-\tau_{\rm ray}\right) d\Omega_{\rm v} dV
\end{equation}

\noindent where $d\dot{n}_{\rm sc}/d\Omega_{\rm v}$ is the number of scatterings within time $dt$ in the volume element $dV$ for which the outgoing photon direction points within solid angle $d\Omega_{\rm v}$ towards the observer. Generally, $d\dot{n}_{\rm sc}/d\Omega_{\rm v}$ and $\tau_{\rm ray}$ are functions of the viewing angle. Due to relativistic beaming, it is convenient to express $d\dot{n}_{\rm sc}/d\Omega_{\rm v}$ in terms of comoving frame properties. From here on, primed quantities are evaluated in the local frame comoving with the outflow. The number of scatterings within the volume element per unit time is Lorentz invariant, since $dV^\prime = \Gamma dV$, $dt^\prime = \Gamma^{-1} dt$ and the total number of scatterings only involves counting. The solid angle transforms as $d\Omega_{\rm v} = \mathcal{D}^{-2} d\Omega^\prime_{\rm v}$ \citep{RybLig:1979}, hence we write $d\dot{n}_{\rm sc}/d\Omega_{\rm v} = \mathcal{D}^2 d\dot{n}^\prime_{\rm sc}/d\Omega^\prime_{\rm v}$. The volume element is $dV = r^2 d\Omega dr$, where $r$ is the radial coordinate and $d\Omega$ is a solid angle element normal to the radial direction. Since the observed photon flux at luminosity distance $d_{\rm L}$ is $(1/d^2_{\rm L}) d\dot{N}_{\rm esc} / d\Omega_{\rm v}$, we write the expression for the energy spectrum,

\begin{equation}
F_E^{\rm ob} (\theta_{\rm v}) = \iint \, \frac{r^2}{d_{\rm L}^2} \mathcal{D}^2 \frac{d\dot{n}^\prime_{\rm sc}}{d\Omega^\prime_{\rm v}} \exp\left(-\tau_{\rm ray}\right) \left\{E \frac{dP}{dE} \right\} d\Omega dr,
\label{eq:short spectrum}
\end{equation}

\noindent where $dP/dE$ is the probability density distribution describing the normalized, local photon spectrum within volume element $dV$ and the integration is over the entire volume.

We now seek the probability for a photon within volume element $dV$ to have a lab frame energy between $E$ and $E + dE$. Under the assumption that photons at $r \ll R_{\rm dcp}$ have a comoving Planck distribution with temperature $T^\prime$, the probability of a photon to have energy between $E^\prime$ and $E^\prime+dE^\prime$ is $dP \propto (B_{\nu^\prime}(T^\prime)/E^\prime) dE^\prime$, where $B_{\nu^\prime}(T^\prime)$ is the Planck spectrum with temperature $T^\prime$. Normalization gives $\int_0^\infty {E^\prime}^2 / [\exp(E^\prime/kT^\prime) - 1] dE^\prime = (kT^\prime)^3 \zeta(3) \approx 2.40 (kT^\prime)^3$ where $\zeta$ is the Riemann zeta function. The observed photon energy and plasma temperature are $E = \mathcal{D} E^\prime$ and $T^{\rm ob} = \mathcal{D} T^\prime$ respectively, and therefore

\begin{equation}
\frac{dP}{dE} = \frac{1}{2.40 (kT^{\rm ob})^3} \frac{E^2}{\exp(E/kT^{\rm ob}) - 1},
\label{eq:norm blackbody}
\end{equation}

\noindent where $k$ is the Boltzmann constant. From Eq. \ref{eq:norm blackbody}, $E(dP/dE) \propto B_\nu(T^{\rm ob})$ within volume element $dV$. As the integration is  over the entire volume, it follows that the observed spectrum is a superposition of blackbodies.

%%%%%%%%%%%%%%%%%%%%%%%%%%%%%%%%%%%%%%%%%%%%%%%%
%%%%%%%%%%%%%%%%%%%%%%%%%%%%%%%%%%%%%%%%%%%%%%%%
%%%%%%%%%%%%%%%%%%%%%%%%%%%%%%%%%%%%%%%%%%%%%%%%
%%%%%%%%%%%%%%%%%%%%%%%%%%%%%%%%%%%%%%%%%%%%%%%%
%%%%%%%%%%%%%%%%%%%%%%%%%%%%%%%%%%%%%%%%%%%%%%%%

\subsection{Angle dependent jet properties}
\label{subsect:angle dependent jet properties}

We summarize the jet interaction with the surrounding environment as an angle dependent mass outflow rate per solid angle, $d\dot{M}/d\Omega = d\dot{M}(\theta)/d\Omega$, an angle dependent photon emission rate per solid angle, $d\dot{N}_\gamma/d\Omega = d\dot{N}_\gamma(\theta)/d\Omega$ and an angle dependent Lorentz factor, $\Gamma = \Gamma(\theta)$, all symmetric around the jet axis. We further assume the Lorentz factor and mass outflow rates to be independent of radius. We omit writing angular dependences in the equations below for clarity. Assuming isotropic scattering in the comoving frame,

\begin{equation}
\frac{d\dot{n}^\prime_{\rm sc}}{d\Omega^\prime_{\rm v}} = \frac{\sigma_{\rm T} c n_{\rm e}^\prime n_\gamma^\prime}{4 \pi}
\label{eq:scattering density}
\end{equation}

\noindent where $c$ is the speed of light. The comoving electron and photon number densities are given by

\begin{equation}
n^\prime_{\rm e} = \frac{1}{r^2 m_{\rm p} c \beta \Gamma} \frac{d\dot{M}}{d\Omega}
\label{eq:electron density}
\end{equation}

\noindent and

\begin{equation}
n^\prime_\gamma = \frac{1}{r^2 c \Gamma} \frac{d\dot{N}_\gamma}{d\Omega}
\label{eq:photon density}
\end{equation}

\noindent respectively, where $m_{\rm p}$ is the proton mass. Using Eq. \ref{eq:electron density}, the decoupling radius can be written as

\begin{equation}
R_{\rm dcp} = \frac{1}{(1 + \beta)\beta\Gamma^2} \frac{\sigma_{\rm T}}{m_{\rm p} c} \frac{d\dot{M}}{d\Omega}
\label{eq:decoupling radius}
\end{equation}

\noindent where use was made of the assumed radial independence of $\Gamma$ and $d\dot{M}/d\Omega$.

The photospheric radius is calculated as follows. Using the fact that $r \sin\theta_{\rm LOS}$ is constant along a ray parallel to the LOS, $dz = -r^2 (r \sin\theta_{\rm LOS})^{-1} d\theta_{\rm LOS}$. Using this change of variables, we write $\tau_{\rm ray} = (\sigma_{\rm T}/m_{\rm p} cr \sin\theta_{\rm LOS}) \int_0^{\theta_{\rm LOS}} [(1-\beta\cos\theta_{\rm LOS})/\beta] (d\dot{M}/d\Omega) d\theta_{\rm LOS}$, from which we obtain

\begin{equation}
R_{\rm ph} = \frac{\sigma_{\rm T}}{m_{\rm p} c} \frac{1}{\sin\theta_{\rm LOS}} \int\limits_0^{\theta_{\rm LOS}} \frac{1 - \beta\cos\tilde{\theta}_{\rm LOS}}{\beta} \frac{d\dot{M}}{d\Omega} d\tilde{\theta}_{\rm LOS}.
\label{eq:photospheric radius}
\end{equation}

\noindent We thus find that for non-zero viewing angles, the photospheric radius is a function of two variables ($R_{\rm ph} = R_{\rm ph}(\theta_{\rm LOS}, \phi_{\rm LOS})$). Since $\tau_{\rm ray} \propto r^{-1}$, we write $\tau_{\rm ray} = R_{\rm ph}/r$. Using Eqs. \ref{eq:short spectrum}, \ref{eq:scattering density}, \ref{eq:electron density}, \ref{eq:photon density} and \ref{eq:decoupling radius} we obtain the observed spectrum,

\begin{eqnarray}
F_E^{\rm ob}  (\theta_{\rm v}) & = & \frac{1}{4 \pi d_{\rm L}^2} \iint \! \, (1+\beta) \mathcal{D}^2 \frac{d\dot{N}_\gamma}{d\Omega} \times \nonumber \\
             &   & \frac{R_{\rm dcp}}{r^2} \exp\left(-\frac{R_{\rm ph}}{r}\right) \left\{ E \frac{dP}{dE} \right\} d\Omega dr.
\label{eq:long spectrum}
\end{eqnarray}

Eq. \ref{eq:long spectrum} (with use of Eqs. \ref{eq:norm blackbody}, \ref{eq:decoupling radius} and \ref{eq:photospheric radius}) provides the energy flux for a given profile. It can be solved once $\Gamma(\theta)$, $d\dot{M}(\theta)/d\Omega$, $d\dot{N}_\gamma(\theta)/d\Omega$ and $\theta_{\rm v}$ are known. In solving Eq. \ref{eq:long spectrum} we consider non-dissipative GRB fireball dynamics \citep{Mes:2006} as a way to relate the angular jet properties. A similar treatment can be used in analyzing other outflow types. For simplicity, we assume an angle independent luminosity,\footnote{As shown below, the part of the spectrum expected to be observed in a prompt GRB is formed by photons making their last scattering at approximately $\theta \lesssim 5/\Gamma_{\rm 0}$. For model JA in \citet{ZhaWooMac:2003}, close to the largest radii of the simulation the Lorentz factor is $\Gamma_{\rm 0} \approx 140$ up to $\theta \approx 0.02$, while $dL/d\Omega \approx const$ up to $\theta \approx 0.05 \approx 7/\Gamma_{\rm 0}$ (as seen in the top panels of Figures 8 and 9 in \citet{ZhaWooMac:2003}). Therefore, $dL/d\Omega = const$ is a good approximation for similar outflows.} $dL/d\Omega = L/4\pi$ where $L$ is the total outflow luminosity. The photon emission rate from the base of the outflow, $r=r_{\rm 0}$, is therefore also independent of angle, $d\dot{N}_\gamma/d\Omega = \dot{N}_\gamma/4\pi$ and $\dot{N}_\gamma = L/2.7 \, kT_{\rm 0}$ where $T_{\rm 0} = (L/4 \pi r_{\rm 0}^2 a c)^{1/4}$ is the temperature at the base of the outflow and $a$ is the radiation constant. Any baryons and associated electrons present are rapidly accelerated by the intense photon field. As the outflow Lorentz factor grows, the angular separation required for causal contact between different parts of the outflow decreases. Therefore, we now make the assumption that a fluid element moving in a direction within solid angle $d\Omega$ evolves separately from the rest of the outflow, as if expanding spherically. The saturated value of the Lorentz factor is equal to the dimensionless entropy, $\eta \equiv dL/d\dot{M} c^2$, and so

\begin{equation}
\Gamma = \frac{L}{4 \pi c^2 d\dot{M}/d\Omega}.
\label{eq:saturated gamma}
\end{equation}

We choose to define $\Gamma = \Gamma(\theta)$ as given by Eq. \ref{eq:gamma profile}, and so $d\dot{M}(\theta)/d\Omega$ follows from Eq. \ref{eq:saturated gamma}. Due to adiabatic energy losses the comoving temperature above saturation decreases as $T^\prime = T_{\rm 0} (r_{\rm 0} / r_{\rm s}) (r_{\rm s}/r)^{2/3}$, where $r_{\rm s} = \eta r_{\rm 0}$ is the saturation radius.

%%%%%%%%%%%%%%%%%%%%%%%%%%%%%%%%%%%%%%%%%%%%%%%%
%%%%%%%%%%%%%%%%%%%%%%%%%%%%%%%%%%%%%%%%%%%%%%%%
%%%%%%%%%%%%%%%%%%%%%%%%%%%%%%%%%%%%%%%%%%%%%%%%
%%%%%%%%%%%%%%%%%%%%%%%%%%%%%%%%%%%%%%%%%%%%%%%%
%%%%%%%%%%%%%%%%%%%%%%%%%%%%%%%%%%%%%%%%%%%%%%%%
%%%%%%%%%%%%%%%%%%%%%%%%%%%%%%%%%%%%%%%%%%%%%%%%
%%%%%%%%%%%%%%%%%%%%%%%%%%%%%%%%%%%%%%%%%%%%%%%%
%%%%%%%%%%%%%%%%%%%%%%%%%%%%%%%%%%%%%%%%%%%%%%%%
%%%%%%%%%%%%%%%%%%%%%%%%%%%%%%%%%%%%%%%%%%%%%%%%
%%%%%%%%%%%%%%%%%%%%%%%%%%%%%%%%%%%%%%%%%%%%%%%%

\section{Contributions from different jet regions to the observed spectrum}
\label{sect:the observed spectrum}

The observed spectrum is obtained by solving Eq. \ref{eq:long spectrum}, namely integrating the emissivity over the entire volume. While analytical integration is difficult for the different jet profiles, numerical integration is straight forward. In \S \ref{sect:the observed spectrum} - \ref{subsect:envelope component} we assume the outflow to be viewed head-on. Therefore the radiative contributions from the different parts of the jet are symmetric around the LOS and Eq. \ref{eq:long spectrum} is readily integrated over azimuthal angle. Non-zero viewing angles are discussed in \S \ref{subsect:non-zero viewing angles} and further explored numerically in \S \ref{sect:results}.

\begin{figure}
\includegraphics[width=\linewidth]{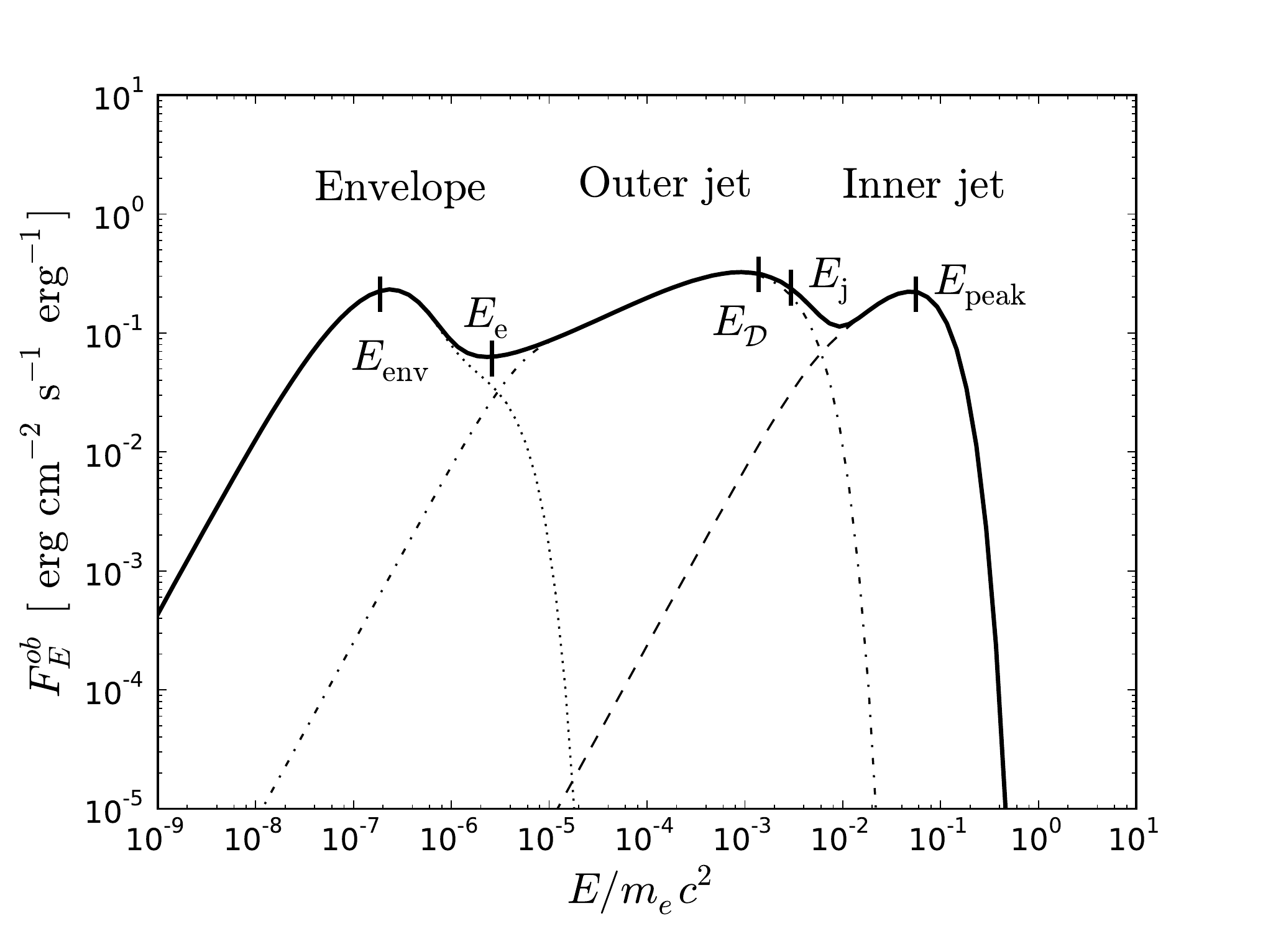}
\caption{The observed spectrum from a relativistic, optically thick outflow obtained by numerical integration of Eq. \ref{eq:long spectrum}. Zero viewing angle is assumed. Separate integration of the contributions from the inner jet, outer jet and envelope is shown with dashed, dash-dotted and dotted lines respectively. Photon energies associated with characteristic outflow angles are indicated by small vertical lines. For this figure $\Gamma_{\rm 0} = 100$, $\theta_{\rm j} = 0.03$ and $p=4$. A total outflow luminosity of $L = 10^{52} \, {\rm erg \, s^{-1}}$ and base outflow radius $r_{\rm 0} = 10^8$ are used.}
\label{fig:general spectrum}
\end{figure}

In Figure \ref{fig:general spectrum} we present the numerically integrated spectrum where we have used the Lorentz factor profile defined in Eq. \ref{eq:gamma profile}. For this figure we use the parameters $\Gamma_{\rm 0} = 100$, $\theta_{\rm j} = 0.03$ and $p = 4$, as well as $\theta_{\rm v} = 0$. The spectrum has a wavy shape, with a slope of roughly $F_E \propto E^0$ over approximately 6 orders of magnitude in energy, which does not resemble the Planck spectrum. We identify several characteristic photon energies in the spectrum ($E_{\rm peak}, \, E_{\rm j}, \, E_\mathcal{D}, \, E_{\rm e}$ and $E_{\rm env}$). Analytical expressions for these energies are given below.

The spectrum can be understood as a combination of components originating from different jet regions. Spectra integrated separately between the angular limits for the inner jet, outer jet and envelope regions are shown in Figure \ref{fig:general spectrum} as dashed, dash-dotted and dotted lines respectively. The advantage of considering the emission from different jet regions separately is that it allows for analytical expressions of key spectral features.

\begin{figure}
\includegraphics[width=\linewidth]{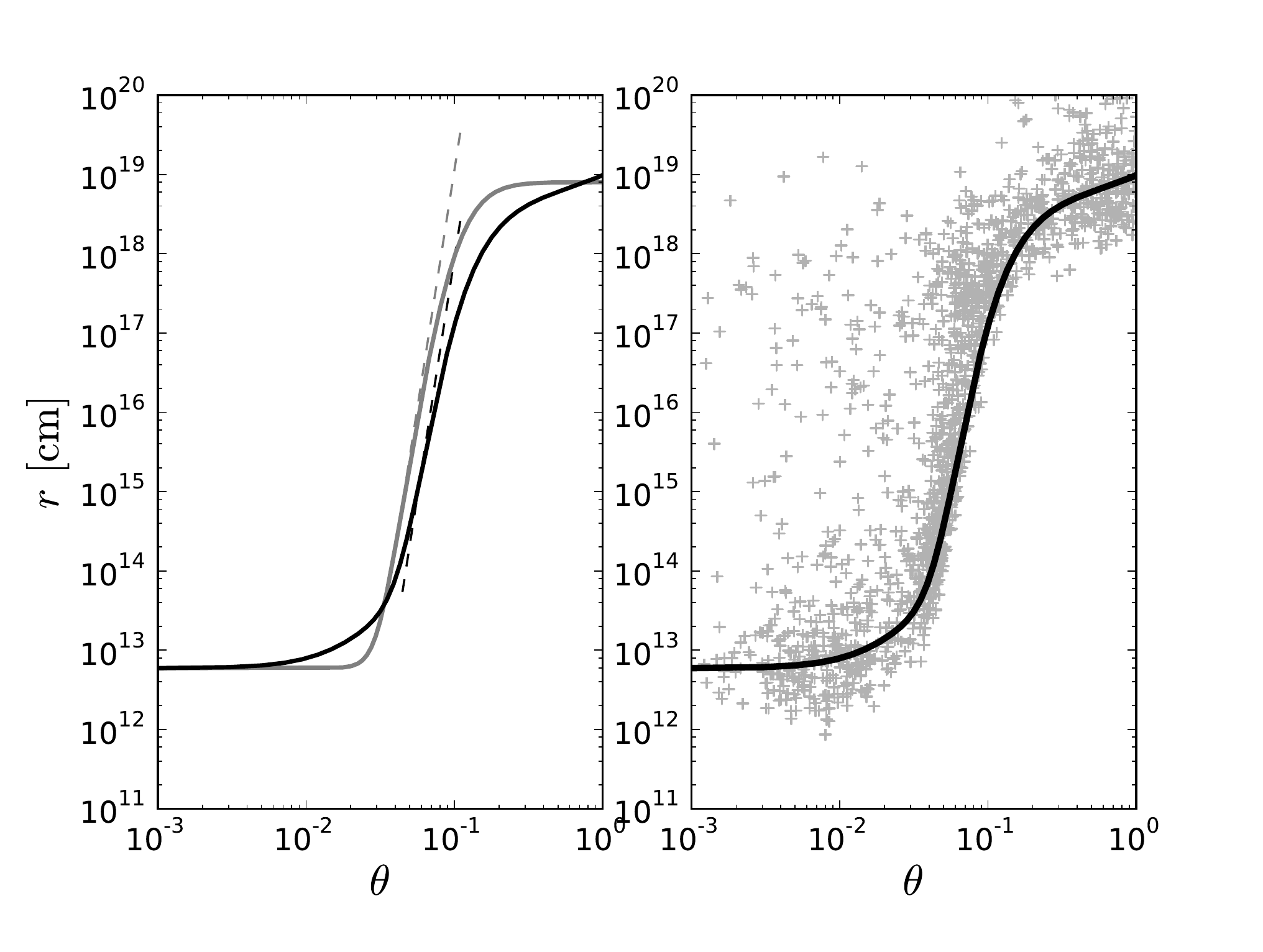}
\caption{The left panel shows the photospheric radius ($R_{\rm ph}$, black) and the photon decoupling radius ($R_{\rm dcp}$, gray) as functions of angle. The dashed lines are the power law approximations used in \S \ref{subsect:outer jet component}. The right panel shows the same photospheric radius (black), and Monte Carlo simulations of the photon last scattering positions before reaching the observer (gray) for $\theta_{\rm v} = 0$ ($0.000 \leq \theta_{\rm v} \leq 0.006$). The Lorentz factor profile $\Gamma_{\rm 0} = 100$, $\theta_{\rm j} = 0.03$, $p = 4$ was used along with $\theta_{\rm v} = 0$. A total outflow luminosity of $L = 10^{52} \, {\rm erg \, s^{-1}}$ was assumed in calculating the radii.}
\label{fig:radii_approx}
\end{figure}

While the integration is over the entire volume, in order to explain the spectrum one can define characteristic photon energies as a function of angle to the jet axis. This is done in the following way. Consider only photons that make their last scattering between angles $\theta$ and $\theta + d\theta$. The distribution of last scattering radii spans a large range of radii (see Figure \ref{fig:radii_approx}, right panel). At $r \ll R_{\rm ph}$ the flux is attenuated by the large optical depth, while at $r \gg R_{\rm ph}$ the probability of scattering is low due to the optical depth being much smaller than unity. Therefore, the characteristic radius of the radial distribution is $R_{\rm ph}(\theta)$. The distribution of observed energies of photons that make their last scattering between angles $\theta$ and $\theta + d\theta$ peaks at $E \approx 2.7 kT^{\rm ob} = 2.7 \, \mathcal{D}(\theta, \Gamma) \, kT^\prime(\theta, R_{\rm ph}(\theta))$. Therefore, one may consider $E(\theta) = 2.7 kT^{\rm ob}(\theta, R_{\rm ph}(\theta))$ to be the characteristic energy of photons making their last scattering between angles $\theta$ and $\theta + d\theta$. Typically, the Doppler boost decreases with angle while the photospheric radius increases with angle.\footnote{The exact angular behaviour of the Doppler boost and the photospheric radius depends on the angular dependence of the outflow parameters.} Therefore, the characteristic photon energy is expected to decrease with angle. Below we use this approximation to obtain the characteristic photon energies shown in Figure \ref{fig:general spectrum}. For $\theta_{\rm v} > 0$, this discussion has to be generalized to include azimuthal angle.

\subsection{Inner jet component}
\label{subsect:inner jet component}

Emission from the inner jet region ($0 \leq \theta \leq \theta_{\rm j}$) is shown with a dashed line in Figure \ref{fig:general spectrum}. The most energetic part of the spectrum originates from this region due to the high Doppler boost and low photospheric radius (see Figure \ref{fig:radii_approx}). Within this region the Lorentz factor is approximately constant. Therefore the observed spectrum from the inner jet region is similar to that of an optically thick, spherically symmetric wind with $\Gamma \approx \Gamma_{\rm 0}$ when only considering the emissivity within $0 \leq \theta \leq \theta_{\rm j}$. The Doppler boost, $\mathcal{D} \approx 2\Gamma/(1 + \Gamma^2 \theta^2)$, and the photospheric radius for a spherically symmetric wind, $R_{\rm ph} \propto (1 + \Gamma^2 \theta^2/3)$ \citep{Pee:2008}, are approximately constant for angles up to $\theta \approx 1/\Gamma_{\rm 0}$, and so the characteristic energy of photons making their last scattering within $0 \leq \theta \leq 1/\Gamma_{\rm 0}$ is approximately constant and is equal to $E_{\rm peak}$.

The observed temperature of photons originating from angles larger than $1/\Gamma_{\rm 0}$ decreases with angle due to the increasing photospheric radius and decreasing Doppler boost. For $1/\Gamma_{\rm 0} < \theta < \theta_{\rm j}$, $R_{\rm ph}/R_{\rm dcp} \propto (1 + \Gamma_{\rm 0}^2 \theta^2/3)$ (see Figure \ref{fig:radii_approx}).\footnote{This follows from $R_{\rm ph} = R_{\rm dcp}$ at $\theta = \theta_{\rm LOS} = 0$, while $R_{\rm dcp} = const$ and $R_{\rm ph} \propto (1 + \Gamma^2\theta^2/3)$ for a spherically symmetric outflow, which is a good approximation for $\theta < \theta_{\rm j}$.} Therefore, the approximation of thermal equillibrium close to the photosphere becomes less accurate with increasing angle (within the inner jet region). Since the characteristic photon energy is a monotonically decreasing function of angle for a spherically symmetric outflow, the approximation of thermal equillibrium becomes less accurate with decreasing observed photon energy. By comparing numerically integrated spectra to the full Monte Carlo simulations we find that for $E \gtrsim 10^{-2} E_{\rm peak}$ the assumption of thermal equillibrium can be used.

The superposition of comoving spectra causes softening of the Rayleigh-Jeans part of the observed spectrum from the inner jet. Through simulations we find that the spectral index at $E = 10^{-2} E_{\rm peak}$ from a spherically symmetric outflow is approximately $0.6$ units softer than the Planck spectrum. This agrees with the index found by \citet{Bel:2010} who considered radiative transfer in a spherically symmetric, collisionally heated outflow.

The peak energy of the inner jet component is obtained by considering the Doppler boosted comoving temperature at the photospheric radius along the LOS. It is the same as that from a spherically symmetric fireball with $\Gamma = \Gamma_{\rm 0}$, for which $\mathcal{D} \approx 2\Gamma_{\rm 0}$, $T^\prime = (T_{\rm 0}/\Gamma_{\rm 0})(r_{\rm s}/r)^{2/3}$ and $R_{\rm ph} = L\sigma_{\rm T}/8 \pi m_{\rm p} c^3 \Gamma_{\rm 0}^3$ under the condition $R_{\rm ph} > r_{\rm s}$ (e.g. \citealt{Mes:2006});

\begin{eqnarray}
E_{\rm peak} & = & 5.4 \times k\left(\frac{L}{4 \pi r_{\rm 0}^2 a c} \right)^{1/4} \left(\frac{8 \pi m_{\rm p} c^3 \Gamma_{\rm 0}^4}{r_{\rm 0}} \right)^{2/3} \nonumber \\
             & = & 540 \left(\frac{L}{10^{52}}\right)^{-5/12} \left(\frac{r_{\rm 0}}{10^8}\right)^{1/6} \left(\frac{\Gamma_{\rm 0}}{300}\right)^{8/3} {\rm keV}.
\label{eq:E_peak}
\end{eqnarray}

\noindent Here and below we use cgs units. All characteristic energies below are presented in units of $E_{\rm peak}$. At angles smaller than $\theta_{\rm j}$ the Lorentz factor is approximately constant. Therefore, the observed spectrum is approximately that of a spherical wind for $E > E_{\rm j} \equiv E(\theta_{\rm j})$.\footnote{In practice, the spectrum starts deviating at slightly higher energies since the Lorentz factor is not exactly constant within the inner jet region (see Figure \ref{fig:general spectrum}).} Under the assumption of thermal equillibrium at $R_{\rm ph}(\theta_{\rm j})$ we obtain the lower limit

\begin{equation}
E_{\rm j}/E_{\rm peak} \geq \left(1 + \frac{\Gamma_{\rm 0}^2 \theta_{\rm j}^2}{2}\right)^{-1}\left(1 + \frac{\Gamma_{\rm 0}^2 \theta_{\rm j}^2}{3}\right)^{-2/3}.
\label{eq:E_j}
\end{equation}

\noindent where the approximations $\sin\theta_{\rm j} \approx \theta_{\rm j}$ and $\beta_{\rm j} \approx 1 - 1/2\Gamma_{\rm j}^2$ were used. This expression simplifies to $E_{\rm j}/E_{\rm peak} \gtrsim 4.2 (\Gamma_{\rm 0} \theta_{\rm j})^{-10/3}$ for $\theta_{\rm j} > 1/\Gamma_{\rm 0}$. We thus conclude that observations of spectra down to two orders of magnitude below the peak will reveal deviations from the spherically symmetric scenario if $\theta_{\rm j} \lesssim 5/\Gamma_{\rm 0}$.

\subsection{Outer jet component}
\label{subsect:outer jet component}

The spectrum from the outer jet region ($\theta_{\rm j} \leq \theta \leq \theta_{\rm e}$) is shown with a dash-dotted line in Figure \ref{fig:general spectrum}. It forms an approximate power law between two limiting energies, $E_{\rm e}$ and $E_\mathcal{D}$. We derive analytical expressions for the power law index and the upper and lower energy limits below. No simple analytical expression exists for the spectrum within the remaining energy range $E_\mathcal{D} < E < E_{\rm j}$. However, this energy range is typically small for the parameter space considered in this work.

To simplify the calculation, we approximate the photon energy distribution as a delta function centered on the average observed photon energy within volume element $dV$: $dP/dE = \delta(E - 2.7 \, kT^{\rm ob})$. This approach is validated by the numerical integration of the spectrum. Furthermore, we assume the outflow to be in thermal equillibrium at $r = R_{\rm ph}(\theta)$. This assumption is justified in the outer jet region for $p > 1$ below. We therefore limit the analytical calculation to outflows with $p > 1$. The characteristic photon energy within volume element $dV$ is $2.7 \, kT^{\rm ob} = (2.7 \, \mathcal{D} \, kT_{\rm 0}/\Gamma)(r_{\rm s}/r)^{2/3}$. The delta function variable is changed to $r$ using the relation $\delta(E) = \lvert dE/dr \rvert^{-1} \, \delta(r)$,

\begin{equation}
E \frac{dP}{dE} = E \delta\left(E - 2.7kT^{ob}\right) = \frac{3}{2} r \delta\left(r - r_{\rm s} \left\{\frac{\mathcal{D}}{\Gamma \epsilon}\right\}^{3/2}\right),
\label{eq:delta functions}
\end{equation}

\noindent where $\epsilon \equiv E/2.7 kT_{\rm 0}$ is the observed photon energy in units of the average photon energy at the base of the outflow. Using Eq. \ref{eq:delta functions} one carries the radial integral in Eq. \ref{eq:long spectrum} which becomes

\begin{eqnarray}
F_E^{\rm ob} & = & \frac{\dot{N}_\gamma}{4 \pi d_{\rm L}^2} \, \frac{3}{2} \int\limits_0^\pi \! \, \mathcal{D}^2 \frac{R_{\rm dcp}}{\Gamma r_{\rm 0}} \left\{\frac{\Gamma \epsilon}{\mathcal{D}}\right\}^{3/2} \times \nonumber \\
             &   & \exp\left(-\frac{R_{\rm ph}}{\Gamma r_{\rm 0}} \left\{\frac{\Gamma \epsilon}{\mathcal{D}}\right\}^{3/2}\right) \sin\theta d\theta.
\label{eq:spectrum solution 1}
\end{eqnarray}

\noindent where $1 + \beta \approx 2$, $r_{\rm s} = \Gamma r_{\rm 0}$ and $d\dot{N}_\gamma / d\Omega = \dot{N}_\gamma / 4\pi$ were used.

In order to solve the angular integral in Eq. \ref{eq:spectrum solution 1}, $\Gamma$, $\mathcal{D}$, $R_{\rm dcp}$ and $R_{\rm ph}$ are approximated as power laws of $\theta$. In the outer jet region the Lorentz factor is well approximated as

\begin{equation}
\Gamma = \Gamma_{\rm 0} \left(\frac{\theta_{\rm j}}{\theta}\right)^p.
\label{eq:gamma pl}
\end{equation}

\begin{figure}
\includegraphics[width=\linewidth]{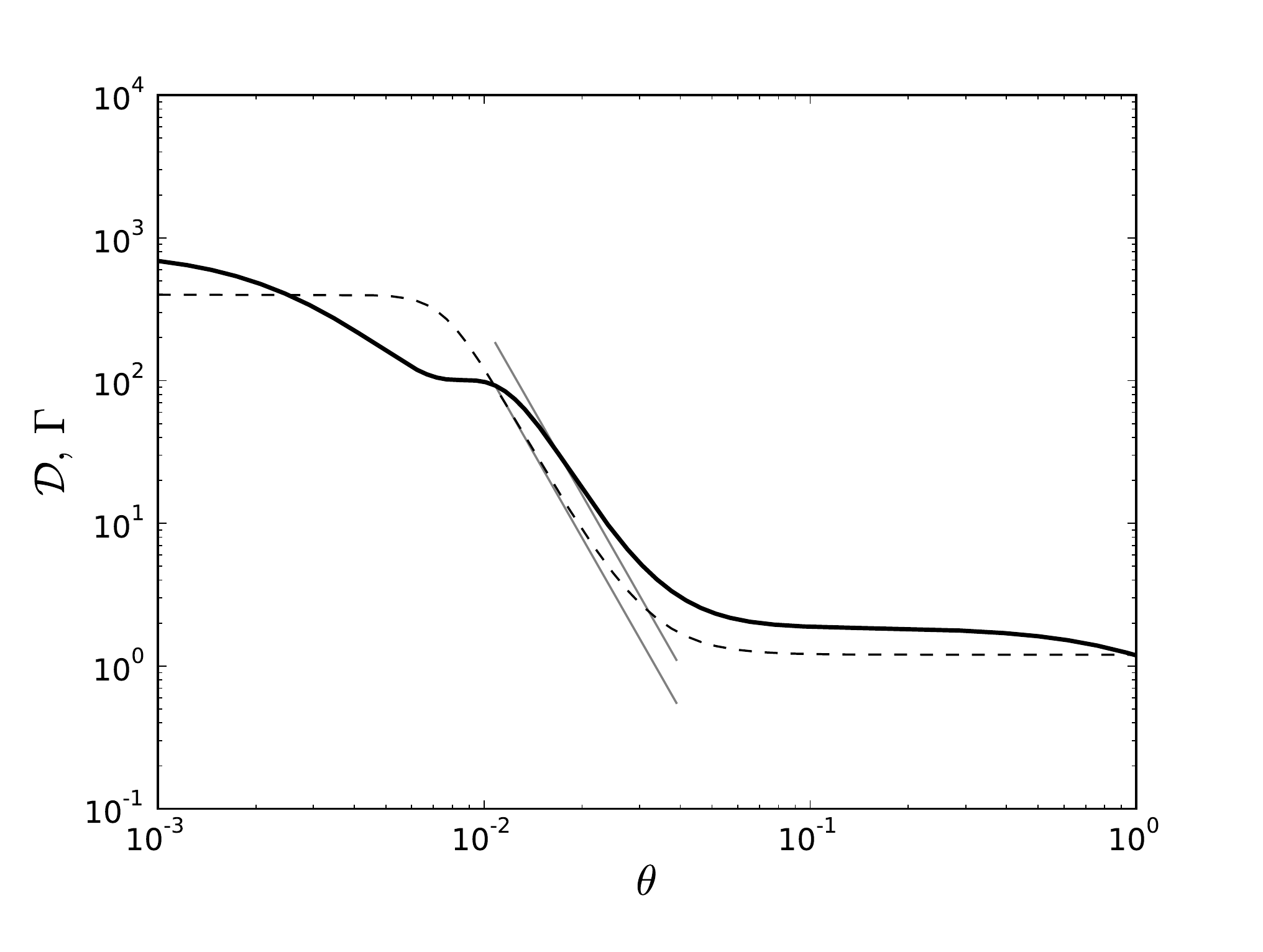}
\caption{The Doppler boost (solid black) and the Lorentz factor (dashed black) as functions of angle to the jet axis. $\theta_{\rm v} = 0$ is assumed. The respective approximations used in \S \ref{subsect:outer jet component} are shown as gray lines. The Doppler boost has a bump at angle $\theta \approx \theta_\mathcal{D}$. The characteristic photon energy associated with $\theta_\mathcal{D}$ is shown in Figure \ref{fig:general spectrum}. A Lorentz factor profile (Eq. \ref{eq:gamma profile}) with parameter values $\Gamma_{\rm 0} = 400$, $\theta_{\rm j} = 3/\Gamma_{\rm 0}$, $p = 4$ was used.}
\label{fig:approximations}
\end{figure}

\noindent For values of $p > 1$, the Doppler boost, $\mathcal{D} \approx 2\Gamma / (1 + \Gamma^2 \theta^2)$ has a break at $\theta_\mathcal{D} \equiv \theta_{\rm j} (\Gamma_{\rm 0} \theta_{\rm j})^{1/(p-1)}$ (see Figure \ref{fig:approximations}).\footnote{The break is located at $\theta_\mathcal{D} \Gamma(\theta_\mathcal{D})  = 1$, and $\Gamma$ is given by Eq. \ref{eq:gamma pl}. For $p = 1$, $\Gamma^2 \theta^2 = const$ and there is no such break.} For angles $\theta_\mathcal{D} < \theta < \theta_{\rm e}$ the desired parameters can be expressed as power laws in $\theta$ to a good approximation, and so we solve for the power law spectrum between these two integration limits. The characteristic photon energies associated with $\theta_\mathcal{D}$ and $\theta_{\rm e}$ gives the upper and lower energy limits $E_\mathcal{D}$ and $E_{\rm e}$. For $\theta_\mathcal{D} < \theta < \theta_{\rm e}$ the Doppler boost is approximately

\begin{equation}
\mathcal{D} = 2\Gamma_{\rm 0} \left(\frac{\theta_{\rm j}}{\theta}\right)^p.
\label{eq:doppler pl}
\end{equation}

\noindent Using the approximate Lorentz factor profile in Eq. \ref{eq:gamma pl} the decoupling radius (Eq. \ref{eq:decoupling radius}) is readily obtained,

\begin{equation}
R_{\rm dcp} = \frac{R_*}{\Gamma_{\rm 0}^3} \left(\frac{\theta}{\theta_{\rm j}}\right)^{3p}
\label{eq:r_rad pl}
\end{equation}

\noindent where $R_{\rm *} \equiv L \sigma_{\rm T} / 8 \pi m_{\rm p} c^3$ and $\beta \approx 1$ were used. Using the small angle approximation $\sin\theta \approx \theta$, the photospheric radius (Eq. \ref{eq:photospheric radius}) at angle $\theta$ is approximately $R_{\rm ph} = (2 R_{\rm *}/\theta) \int_0^\theta \Gamma^{-2} \mathcal{D}^{-1} d\tilde{\theta}$. The integral may be split at angle $\theta_\mathcal{D}$, representing contributions to the optical depth from electrons at larger and smaller angles, respectively. The contribution from within $\theta_\mathcal{D}$ is subdominant for $p > 1$ and $\theta \gg \theta_\mathcal{D}$, and so it is sufficient to consider $\theta_\mathcal{D}$ as the lower angular limit. Using Eqs. \ref{eq:gamma pl} and \ref{eq:doppler pl} the angular integration for the photospheric radius is readily solved,

\begin{equation}
R_{\rm ph} = \frac{R_{\rm *}}{(3p + 1) \Gamma_{\rm 0}^3} \left(\frac{\theta}{\theta_{\rm j}}\right)^{3p},
\label{eq:r_ray pl}
\end{equation}

\noindent which is a factor $3p + 1$ lower than $R_{\rm dcp}$ for all angles $\theta_\mathcal{D} \lesssim \theta \lesssim \theta_{\rm e}$. We therefore conclude that the outflow is in thermal equillibrium at the photosphere within this angular range. The approximated power laws are shown in Figures \ref{fig:radii_approx} (left panel) and \ref{fig:approximations}.

In order to find the photon energy associated with $\theta_\mathcal{D}$, $E_\mathcal{D} \approx 2.7 \, \mathcal{D}(\theta_\mathcal{D}) (kT^\prime / \Gamma(\theta_\mathcal{D}))(r_{\rm s}(\theta_\mathcal{D})/R_{\rm ph}(\theta_\mathcal{D}))^{2/3}$, one first calculates $R_{\rm ph}(\theta_\mathcal{D})$.\footnote{At $\theta_\mathcal{D}$, the electrons at angles $\theta < \theta_\mathcal{D}$ have to be taken into account when calculating the optical depth. By integrating the expression for the photospheric radius (Eq. \ref{eq:photospheric radius}) over angles $\theta < \theta_\mathcal{D}$, one obtains $R_{\rm ph} \approx (R_{\rm *}/\Gamma_{\rm 0}^3) (\Gamma_{\rm 0} \theta_{\rm j})^{-1/(p-1)} \{1 + (\Gamma_{\rm 0} \theta_{\rm j})^2 [1/3 + (p+3)^{-1} (\Gamma_{\rm 0} \theta_{\rm j})^{(p+3)/(p-1)} (1 - (\Gamma_{\rm 0} \theta_{\rm j})^{-(p+3)/(p-1)})] \}$. This simplifies to $R_{\rm ph} \approx (R_{\rm *}/\Gamma_{\rm 0}^3) (p+3)^{-1} (\Gamma_{\rm 0} \theta_{\rm j})^{3p/(p-1)}$ for $\theta_{\rm j} > 1/\Gamma_{\rm 0}$. Note that this expression is different from Eq. \ref{eq:r_ray pl}, which is derived for $\theta \gg \theta_\mathcal{D}$.} Using this approximation and $\mathcal{D}(\theta_\mathcal{D}) = \Gamma(\theta_\mathcal{D})$, we calculate the photon energy associated with $\theta_\mathcal{D}$,

\begin{equation}
E_\mathcal{D}/E_{\rm peak} = \frac{\left(p+3\right)^{2/3}}{2} \left(\Gamma_{\rm 0} \theta_{\rm j}\right)^{-8p/3(p-1)},
\label{eq:E_D}
\end{equation}

\noindent which is valid for $\theta_{\rm j} > 1/\Gamma_{\rm 0}$. As can be seen from the definition of $\theta_\mathcal{D}$ (below Eq. \ref{eq:gamma pl}), $\theta_\mathcal{D} \approx \theta_{\rm j}$ when $\theta_{\rm j} \approx 1/\Gamma_{\rm 0}$. Therefore $E_\mathcal{D} \approx E_{\rm j} \approx E_{\rm peak}$ when $\theta_{\rm j} \approx 1/\Gamma_{\rm 0}$. Thus, for narrow jets ($\theta_{\rm j} \approx 1/\Gamma_{\rm 0}$) the upper limiting energy to the power law is approximately the same as the peak from the inner jet region. For $\theta_{\rm j} > 1/\Gamma_{\rm 0}$ there is a bump-like spectral feature related to $E_\mathcal{D}$ (see Figure \ref{fig:general spectrum}). If this feature is observed, constrains can be put on the jet properties through Eq. \ref{eq:E_D}.

Inserting Eqs. \ref{eq:gamma pl}, \ref{eq:doppler pl}, \ref{eq:r_rad pl} and \ref{eq:r_ray pl} into Eq. \ref{eq:spectrum solution 1} gives the observed flux within the energy range $E_{\rm e} \ll E \ll E_\mathcal{D}$,

\begin{eqnarray}
F_E^{\rm ob} & = & \frac{\dot{N}_\gamma}{4 \pi d_{\rm L}^2} \, \frac{3}{2} \, \frac{4}{\left(\Gamma_{\rm 0} \theta_{\rm j}^p\right)^2} \frac{R_{\rm *}}{r_{\rm 0}}\left(\frac{\epsilon}{2}\right)^{3/2} \int\limits_{\theta_{\rm e}}^{\theta_\mathcal{D}} \! \theta^{2p + 1} \times \nonumber \\
             &   & \exp\left(-\frac{1}{3p + 1} \frac{1}{(\Gamma_{\rm 0} \theta_{\rm j}^p)^4} \frac{R_{\rm *}}{r_{\rm 0}}\left(\frac{\epsilon}{2}\right)^{3/2} \theta^{4p}\right) d\theta \nonumber \\
             & = & \frac{\dot{N}_\gamma}{4 \pi d_{\rm L}^2} \, \frac{3}{2} \, \frac{\left(3p + 1\right)^{\frac{1}{2}\left(\frac{1}{p} + 1\right)}}{p} \left(\Gamma_{\rm 0} \theta_{\rm j}^p\right)^\frac{2}{p} \left(\frac{R_{\rm *}}{r_{\rm 0}}\right)^{\frac{1}{2}\left(1-\frac{1}{p}\right)} \times \nonumber \\
             &   & \left(\frac{\epsilon}{2}\right)^{\frac{3}{4}(1 - \frac{1}{p})} \int\limits_{x_{\rm min}}^{x_{\rm max}} \! x^{\frac{1}{2}\left(\frac{1}{p} - 1\right)} \exp\left(-x\right) dx
\label{eq:spectrum solution 2}
\end{eqnarray}

\noindent where $x \equiv (3p + 1)^{-1} (\Gamma_{\rm 0} \theta_{\rm j}^p)^{-4} (R_{\rm *}/r_{\rm 0}) (\epsilon/2)^{3/2} \theta^{4p}$, $x_{\rm min} = (3p+1)^{-1} (R_{\rm *}/r_{\rm 0}) (\epsilon/2)^{3/2} (\Gamma_{\rm 0} \theta_{\rm j}^p)^{4/p-1}$ and $x_{\rm max} = (3p+1)^{-1} (R_{\rm *}/r_{\rm 0}) (\epsilon/2)^{3/2} ((1 + \sqrt{2})/\Gamma_{\rm min})^4$.

The integral in Eq. \ref{eq:spectrum solution 2} is approximately constant for a large range of energies. The integrand decreases exponentially for large $x$. When $x_{\rm max} \gg 1$, the upper limit is well approximated as infinity. This condition translates to

\begin{equation}
\epsilon \gg \frac{E_{\rm e}}{2.7 \, kT_{\rm 0}} \equiv 2 \left\{\left(3p+1\right) \frac{r_{\rm 0}}{R_{\rm *}} \left(\frac{\Gamma_{\rm min}}{1 + \sqrt{2}}\right)^4\right\}^{2/3}.
\label{eq:lower energy}
\end{equation}

\noindent In units of the peak energy, one obtains

\begin{equation}
E_{\rm e}/E_{\rm peak} = \frac{(3p+1)^{2/3}}{(\sqrt{2} + 1)^{8/3}} \left(\frac{\Gamma_{\rm min}}{\Gamma_{\rm 0}}\right)^{8/3}.
\label{eq:E_e}
\end{equation}

\noindent Setting the upper limit to infinity, the integral in Eq. \ref{eq:spectrum solution 2} becomes

\begin{equation}
\int\limits_{x_{\rm min}}^\infty \! x^{\frac{1}{2}\left(\frac{1}{p} - 1\right)} \exp\left(-x\right) dx = \Gamma_{\rm inc}\left\{\frac{1}{2}\left(\frac{1}{p} + 1\right), \, x_{\rm min}\right\}
\label{eq:x integral}
\end{equation}

\noindent where $\Gamma_{\rm inc}\{(1/p + 1)/2, \, x_{\rm min}\}$ is the incomplete gamma function, which for $x_{\rm min} \ll 1$ and $p > 1$ is approximately constant.\footnote{For $x_{\rm min} = 0$ the incomplete gamma function equals the complete gamma function, $\Gamma\{(1/p + 1)/2\}$. For $1 < p < 10^2$, $1 < \Gamma\{(1/p + 1)/2\} < 1.76$.} The condition $x_{\rm min} \ll 1$ translates to

\begin{equation}
\epsilon \ll 2 \left\{\left(3p+1\right) \frac{r_{\rm 0}}{R_{\rm *}} \left(\Gamma_{\rm 0} \theta_{\rm j}^p\right)^{-\frac{4}{p-1}}\right\}^{2/3} \approx \frac{E_\mathcal{D}}{kT_{\rm 0}}.
\end{equation}

We thus find that the integral in Eq. \ref{eq:spectrum solution 2} is approximately constant within the energy range $E_{\rm e} \ll E \ll E_\mathcal{D}$. The observed flux in this energy range is a power law approximately satisfying

\begin{equation}
F_E^{\rm ob} \propto E^{\frac{3}{4}\left(1 - \frac{1}{p}\right)}
\label{eq:spectrum solution 4}
\end{equation}

\noindent for $p > 1$. Expressed as the photon index within this energy range, Eq. \ref{eq:spectrum solution 4} implies $\alpha \approx -(1/4)(1 + 3/p)$. Redoing the above calculation for $p = 1$ results in the same power law index, with the modification that the upper limiting energy is set by $E_{\rm j}$, and so Eq. \ref{eq:spectrum solution 4} can be considered valid for $p \geq 1$. We therefore conclude that for a Lorentz factor profile with $p = 1$, $\alpha \approx -1$ while for $p = 2$, $\alpha \approx -0.5$. For an angular profile with $p = 4$ and $\theta_{\rm j} = 3/\Gamma_{\rm 0}$ (as shown in Figure \ref{fig:general spectrum}), $E_\mathcal{D}/E_{\rm peak} = 2.5 \times 10^{-2}$ and $E_{\rm e}/E_{\rm peak} = 4.6 \times 10^{-5}$. For jets with large opening angles ($\theta_{\rm j} \gtrsim 10/\Gamma_{\rm 0}$), the range of energies between $E_\mathcal{D}$ and $E_{\rm e}$ is small and so the outer jet spectrum is not well approximated as a power law.

\subsection{Envelope component}
\label{subsect:envelope component}

The spectrum from the envelope component ($\theta \leq \theta_{\rm e}$) is shown with a dotted line in Figure \ref{fig:general spectrum}. It is characterized by a peak at $E = E_{\rm env}$, where $E_{\rm env}$ is given by the observed temperature of the envelope component.

The Lorentz factor of the envelope is approximately constant, close to unity. Due to the low bulk speed, the photospheric radius of the envelope is very large. Therefore, the comoving photon temperature at transparency is very low. Furthermore, the observed envelope temperature is barely Doppler boosted. Therefore, its spectrum is approximately that of a blackbody with observed temperature equal to the comoving temperature at the radius of transparency. For $\Gamma_{\rm min} = 1.2$, the envelope speed is $\beta_{\rm min} \approx 0.5$, while $\cos\theta_{\rm e} \approx 1$ for most model parameters. Therefore, the Doppler boost is $\mathcal{D} \approx [\Gamma_{\rm min}(1 - \beta_{\rm min})]^{-1} = \Gamma_{\rm min}(1+\beta_{\rm min})$. The photospheric radius for angles larger than $\theta_{\rm e}$ is approximately constant, equal to the decoupling radius since the electrons move at a bulk speed that is only mildly relativistic (see Figure \ref{fig:radii_approx}). Thus we calculate the ratio of the characteristic observed envelope photon energy to the peak photon energy,

\begin{equation}
E_{\rm env}/E_{\rm peak} \approx \left(\frac{1+\beta_{\rm min}}{2}\right)^{5/3} \beta_{\rm min}^{2/3} \left(\frac{\Gamma_{\rm min}}{\Gamma_{\rm 0}}\right)^{8/3}.
\label{eq:E_env}
\end{equation}

\noindent For an angular profile with $\Gamma_{\rm 0} = 100$ and $\Gamma_{\rm min} = 1.2$, $E_{\rm env}/E_{\rm peak} \approx 3.3 \times 10^{-6}$.

\subsection{Non-zero viewing angles}
\label{subsect:non-zero viewing angles}

Due to the complexity of the calculation in this scenario, we give here only a qualitative explanation. Full quantitative treatment is presented in \S \ref{sect:results}. Increasing the viewing angle has two main consequences: decreasing $E_{\rm peak}$ (and therefore also the bolometric flux) and softening the spectrum. Since the characteristic angles for the inner jet are smaller than those for the outer jet, the inner jet spectral component is affected more than the outer jet component by viewing angle variations.

As long as $\theta_{\rm v} \ll \theta_{\rm j}$ the observed spectrum is well approximated by the $\theta_{\rm v} = 0$ solution. For $\theta_{\rm v} > \theta_{\rm j}$, $E_{\rm peak}$ decreases due to the decreasing Doppler boost as well as the increasing photospheric radius. Photons from the outer jet region contribute to a general softening of the spectrum below the peak energy. For $\theta_{\rm v} > \theta_{\rm j}$, the total observed flux decreases with increasing viewing angle, and at some viewing angle (dependent on jet parameters as well as detector characteristics) the flux in non-detectable.

For $p \rightarrow \infty$ the jet profile approaches a top-hat shape. For such jets the peak energy and bolometric flux decreases rapidly with increasing viewing angle (for $\theta_{\rm v} > \theta_{\rm j}$) and the probability of observing the outflow at $\theta_{\rm v} > \theta_{\rm j}$ is low. However, for outflows with a shallow Lorentz factor gradient (small values of $p$), the bolometric flux and peak energy decrease are slower. Therefore the outflow may be observed up to $\theta_{\rm v} \approx few \times \theta_{\rm j}$, depending on detector characteristics and the exact jet profile. Assume $\theta_{\rm v, max}$ to be the largest viewing angle where the outflow is still detectable. The fraction of GRBs observed at $\theta_{\rm v} < \theta_{\rm j}$ is $\Delta\Omega_{\rm j}/\Delta\Omega_{\rm v, max} = \theta_{\rm j}^2 / \theta_{\rm v, max}^2$. For $\theta_{\rm v, max} > \sqrt{2} \times \theta_{\rm j}$, the majority of observed outflows are viewed at $\theta_{\rm v} > \theta_{\rm j}$. Therefore, the consequences of observing the outflow at $\theta_{\rm v} \approx \theta_{\rm j}$ should not be neglected.

%%%%%%%%%%%%%%%%%%%%%%%%%%%%%%%%%%%%%%%%%%%%%%%%
%%%%%%%%%%%%%%%%%%%%%%%%%%%%%%%%%%%%%%%%%%%%%%%%
%%%%%%%%%%%%%%%%%%%%%%%%%%%%%%%%%%%%%%%%%%%%%%%%
%%%%%%%%%%%%%%%%%%%%%%%%%%%%%%%%%%%%%%%%%%%%%%%%
%%%%%%%%%%%%%%%%%%%%%%%%%%%%%%%%%%%%%%%%%%%%%%%%
%%%%%%%%%%%%%%%%%%%%%%%%%%%%%%%%%%%%%%%%%%%%%%%%
%%%%%%%%%%%%%%%%%%%%%%%%%%%%%%%%%%%%%%%%%%%%%%%%
%%%%%%%%%%%%%%%%%%%%%%%%%%%%%%%%%%%%%%%%%%%%%%%%
%%%%%%%%%%%%%%%%%%%%%%%%%%%%%%%%%%%%%%%%%%%%%%%%
%%%%%%%%%%%%%%%%%%%%%%%%%%%%%%%%%%%%%%%%%%%%%%%%

\section{Numerical simulation}
\label{sect:numerical simulations}

In order to validate the analytical model and to explore the importance of full photon propagation history, a numerical code has been developed. The code is based on an earlier code for photon propagation in a spherically symmetric, relativistically expanding plasma \citep{PeeWax:2004, PeeMesRee:2006, Pee:2008}. This code is specially designed to allow for the bulk Lorentz factor, mass outflow rate and photon emission rate to be functions of angle with respect to the jet axis. This is required in the current context as the optical depth between two points in the outflow depends on the local Lorentz factor and electron density. As photons propagate across the jet, photon energy changes due to propagation between regions of different bulk Lorentz factor is inherently considered; therefore the comoving photon spectrum at the last scattering position is not artificially fixed to blackbody. No assumption of thermal equillibrium is made above the photon injection position, as opposed to the analytical model. The full Klein-Nishina cross section is used in the scattering process.

\subsection{Code description}
\label{subsect:code description}

The code tracks photon propagation, starting from a random position in the outflow with $\tau_{\rm rdl} = 20$, ensuring a low probability ($\exp(-20) \approx 2.1 \times 10^{-9}$) for a photon to escape the outflow without being scattered. The lab frame angular coordinates of the initial position are randomly drawn from a uniform distribution. The initial photon propagation direction is randomly chosen in a similar way. The initial comoving photon energy is drawn from a blackbody distribution with temperature equal to the comoving outflow temperature at the photon position; $T^\prime(\theta, r) = T_{\rm 0} (r_{\rm 0}/r_{\rm s}(\theta))(r_{\rm s}(\theta)/r)^{2/3}$.

A photon has a probability $\exp(-\tau)$ to propagate an optical depth $\tau$ before scattering. Thus, an optical depth, $\Delta\tau$, representing the distance between scatterings in the direction of propagation, is drawn from a logarithmic distribution. The optical depth from the photon position to infinity in the photon propagation direction, $\tau_\infty$, is calculated and compared to the drawn optical depth. If $\Delta\tau \geq \tau_\infty$ the photon escapes the outflow, else the distance corresponding to $\Delta\tau$ is calculated, taking into account the angular dependence of the Lorentz factor and electron number density. The new photon scattering position is then considered. The photon four-vector is transformed to the local comoving outflow frame. The scattering electron Lorentz factor is drawn from a Maxwellian distribution with temperature $T^\prime(\theta, r)$ along with a random electron propagation direction. The photon four-vector is transformed to the electron rest frame and the photon undergoes a Compton scattering, changing its energy and direction. The photon four-vector is transformed back to the lab frame, and a new optical depth is drawn. The process is repeated until the photon has escaped the outflow.

After simulating a sufficient number of photons (typically $3 \times 10^7$) the escaping photons are binned in viewing angles. The bin width is dependent on how sensitive the spectrum from a given profile is for viewing angle variations. The bin width used for each spectrum is presented in the respective figure caption. A typical width is $\Delta\theta_{\rm v} \approx 0.005$.

\section{Results from simulation and numerical integration of the spectrum}
\label{sect:results}

\begin{figure}
\includegraphics[width=\linewidth]{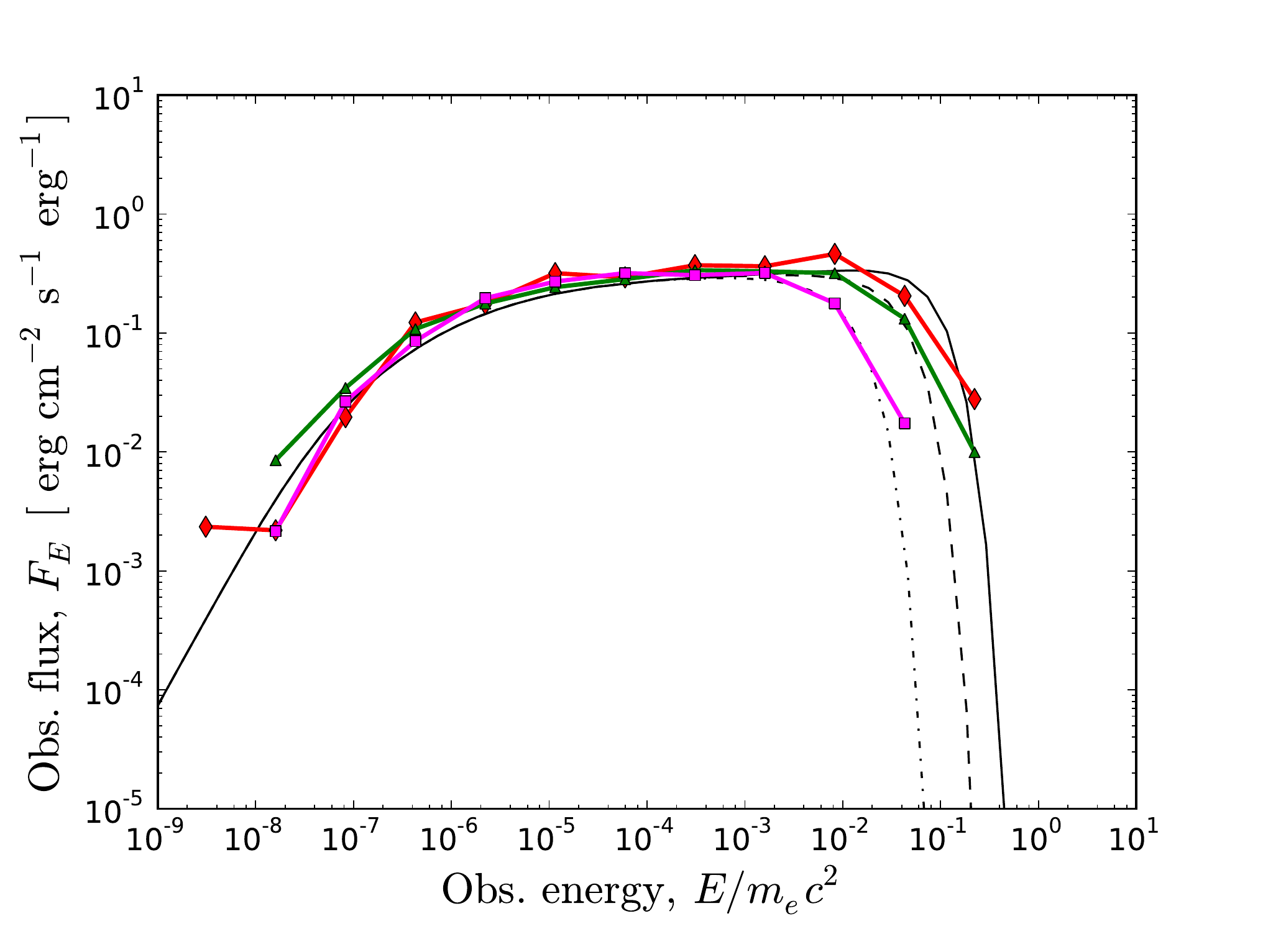}
\caption{Simulated (colored) and numerically integrated (black) spectra for a narrow jet observed at different viewing angles. In this plot, $\Gamma_{\rm 0} = 100$, $\theta_{\rm j} = 1/\Gamma_{\rm 0}$ and $p = 1$. Three different viewing angles are shown, $\theta_{\rm v} = 0$ ($0.000 \leq \theta_{\rm v} \leq 0.0045$, red diamonds and solid black lines), $\theta_{\rm v} = \theta_{\rm j}$ ($0.009 \leq \theta_{\rm v} \leq 0.011$, green triangles and black dashed lines) and $\theta_{\rm v} = 2 \times \theta_{\rm j}$ ($0.019 \leq \theta_{\rm v} \leq 0.020$, magenta squares and dash-dotted lines). After viewing angle binning, the red, green and magenta spectra contain $1002$, $1758$ and $1486$ photons, respectively. The photon index below $E_{\rm peak}$ is $\alpha = -1$ for all viewing angles, two units less than the Rayleigh-Jeans index. For this parameter space region the numerical integration gives an excellent fit to the simulated spectra.}
\label{fig:g100p1j001_vary_v}
\end{figure}

\begin{figure}
\includegraphics[width=\linewidth]{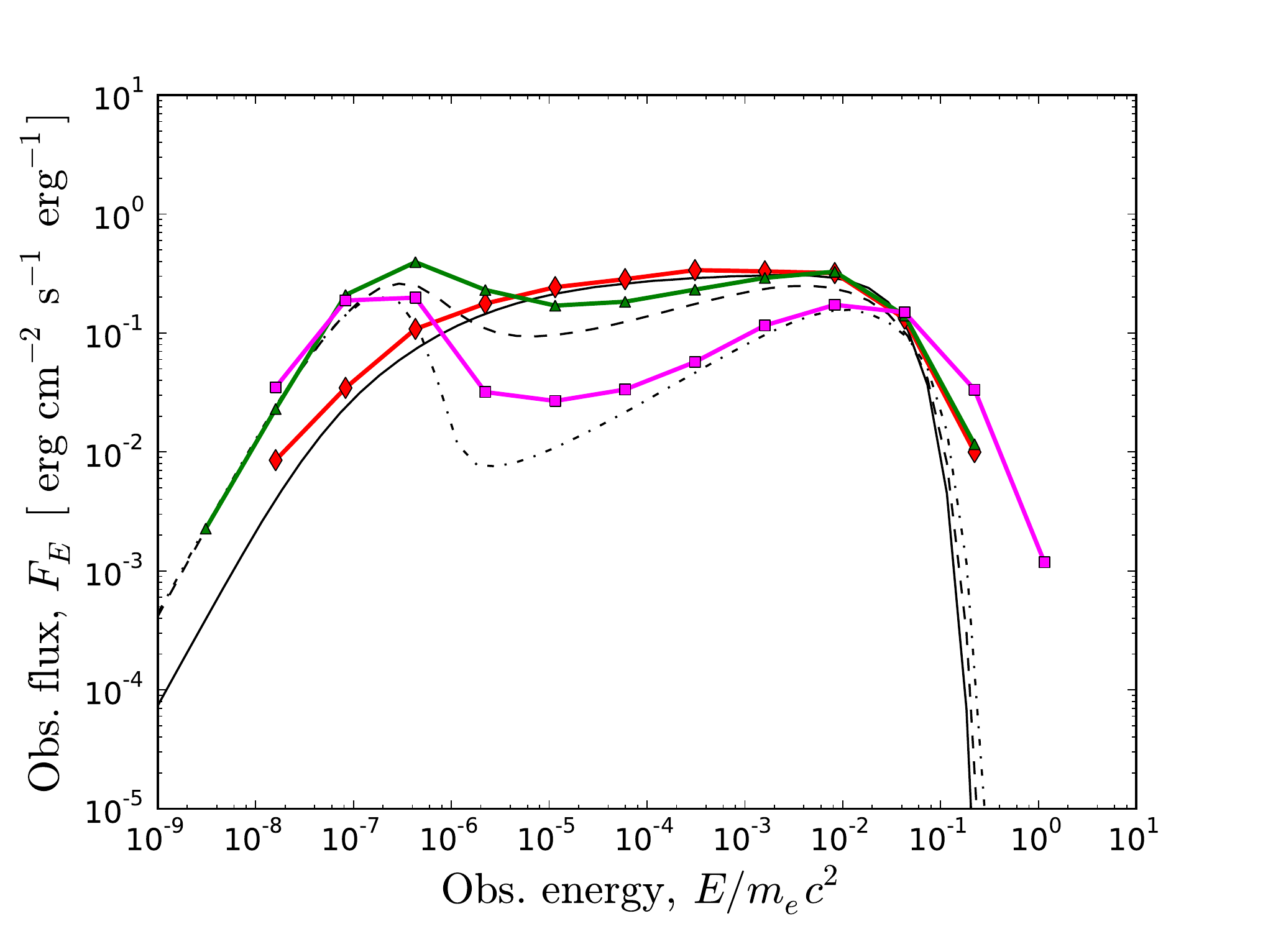}
\caption{Simulated (colored) and numerically integrated (black) spectra for narrow jets with different Lorentz factor gradients. In this plot, $\Gamma_{\rm 0} = 100$, $\theta_{\rm j} = 1/\Gamma_{\rm 0}$ and $\theta_{\rm v} = \theta_{\rm j}$ ($0.009 \leq \theta_{\rm v} \leq 0.011$). Three different values of $p$ are shown, $p = 1$ (red diamonds and solid black lines), $p = 2$ (green triangles and black dashed lines) and $p = 4$ (magenta squares and dash-dotted lines). After viewing angle binning, the red, green and magenta spectra contain $1758$, $1900$ and $891$ photons, respectively. The low energy photon index is close to $\alpha \approx -1$ for all Lorentz factor gradients considered here. For $p=4$, the high energy spectrum does not decay exponentially due to photon diffusion from high angles. See further discussion in the text.}
\label{fig:g100v1j001_vary_p}
\end{figure}

\begin{figure}
\includegraphics[width=\linewidth]{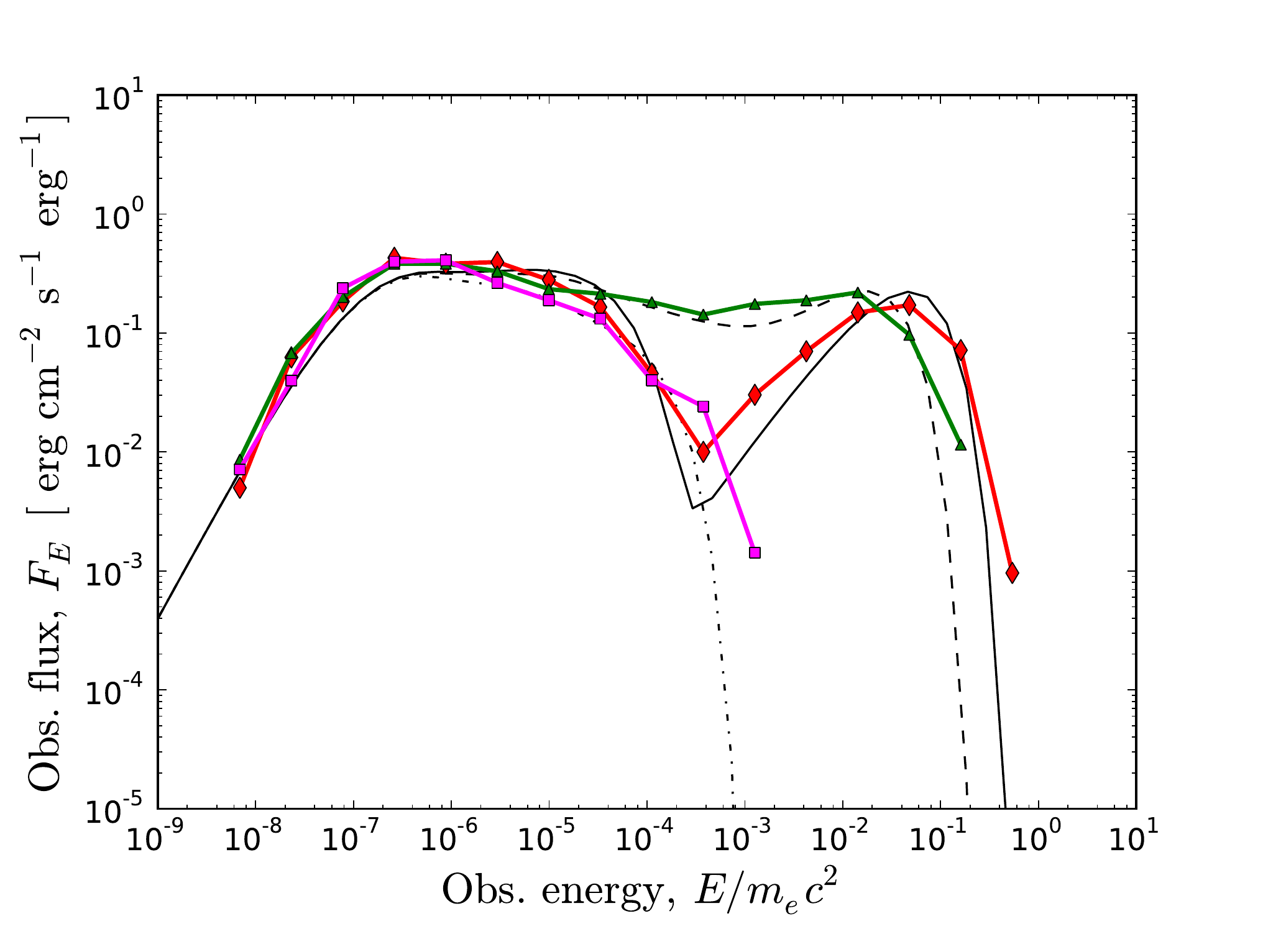}
\caption{Simulated (colored) and numerically integrated (black) spectra for a wide jet observed at different viewing angles. In this plot, $\Gamma_{\rm 0} = 100$, $\theta_{\rm j} = 10/\Gamma_{\rm 0}$ and $p = 4$. Three different viewing angles are shown, $\theta_{\rm v} = 0$ ($0.000 \leq \theta_{\rm v} \leq 0.006$, red diamonds and solid black lines), $\theta_{\rm v} = \theta_{\rm j}$ ($0.009 \leq \theta_{\rm v} \leq 0.011$, green triangles and black dashed lines) and $\theta_{\rm v} = 2 \times \theta_{\rm j}$ ($0.019 \leq \theta_{\rm v} \leq 0.020$, magenta squares and dash-dotted lines). After viewing angle binning, the red, green and magenta spectra contain $1577$, $1826$ and $1112$ photons, respectively. For $\theta_{\rm v} \ll \theta_{\rm j}$, the high energy spectrum resembles that of a spherical wind. For $\theta_{\rm v} = \theta_{\rm j}$, the photon index is $\alpha = -1$, similar to the spectrum from a narrow jet shown in Figure \ref{fig:g100p1j001_vary_v}. Due to the large Lorentz factor gradient ($p = 4$), for $\theta_{\rm v} = 2 \times \theta_{\rm j}$ the peak energy is very low.}
\label{fig:g100p4j010_vary_v}
\end{figure}

\begin{figure}
\includegraphics[width=\linewidth]{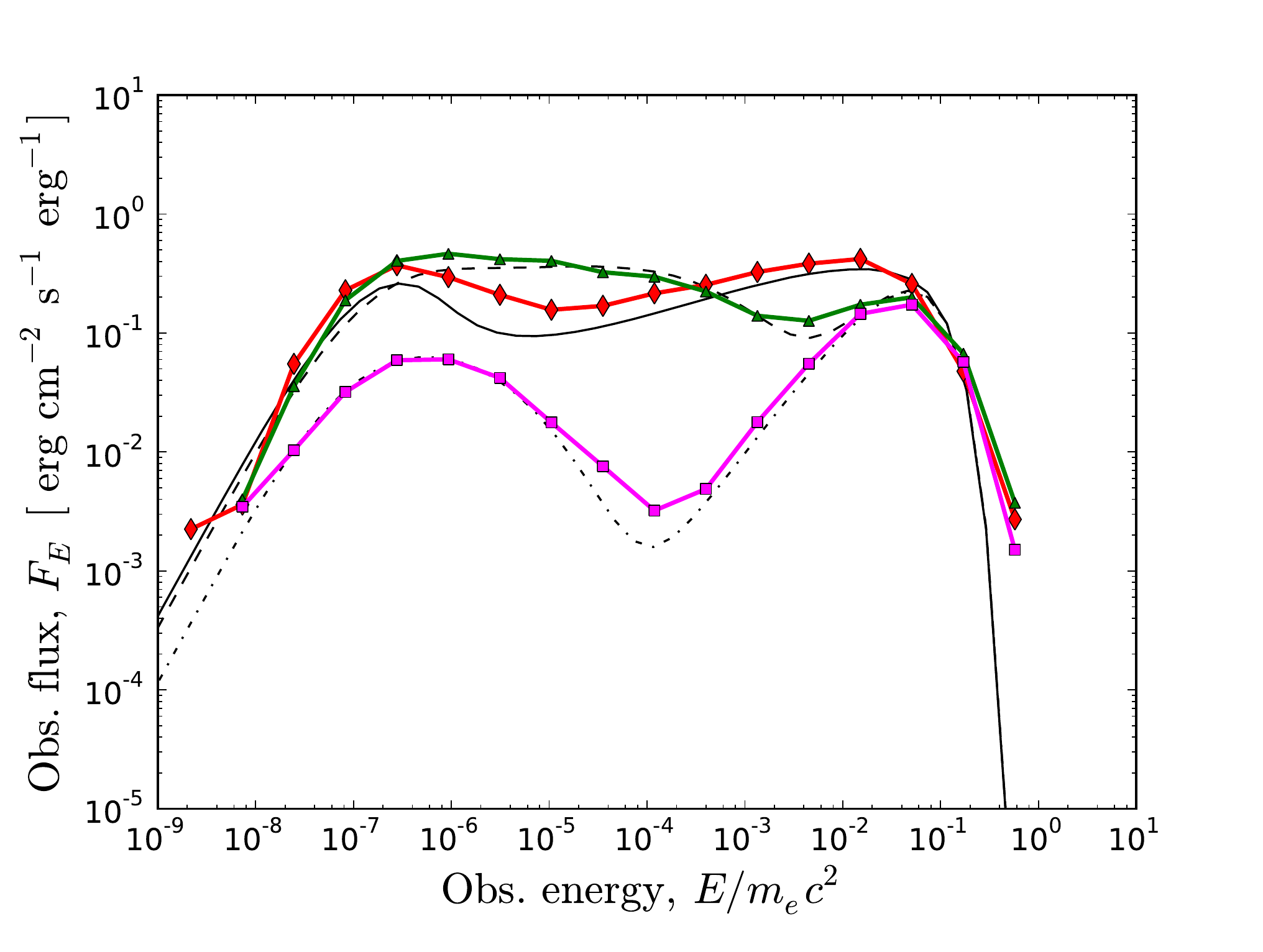}
\caption{Simulated (colored) and numerically integrated (black) spectra for jets with different jet opening angles. In this plot, $\Gamma_{\rm 0} = 100$, $p = 2$ and $\theta_{\rm v} = 0$ ($0.000 \leq \theta_{\rm v} \leq 0.006$ for the red and green spectra, $0.000 \leq \theta_{\rm v} \leq 0.020$ for the magenta spectra). Three different values of $\theta_{\rm j}$ are shown, $\theta_{\rm j} = 1/\Gamma_{\rm 0}$ (red diamonds and solid black lines), $\theta_{\rm j} = 3/\Gamma_{\rm 0}$ (green triangles and black dashed lines) and $\theta_{\rm j} = 10/\Gamma_{\rm 0}$ (magenta squares and dash-dotted lines). After viewing angle binning, the red, green and magenta spectra contain $2163$, $2267$ and $4482$ photons, respectively. The exact spectral shape depends on the value of $p$, however the average spectral index is approximately $-1$ for jet opening angles up to $\theta_{\rm j} \approx few/\Gamma_{\rm 0}$ and values of $p$ up to a few.}
\label{fig:g100p2v0_vary_j}
\end{figure}

\begin{figure}
\includegraphics[width=\linewidth]{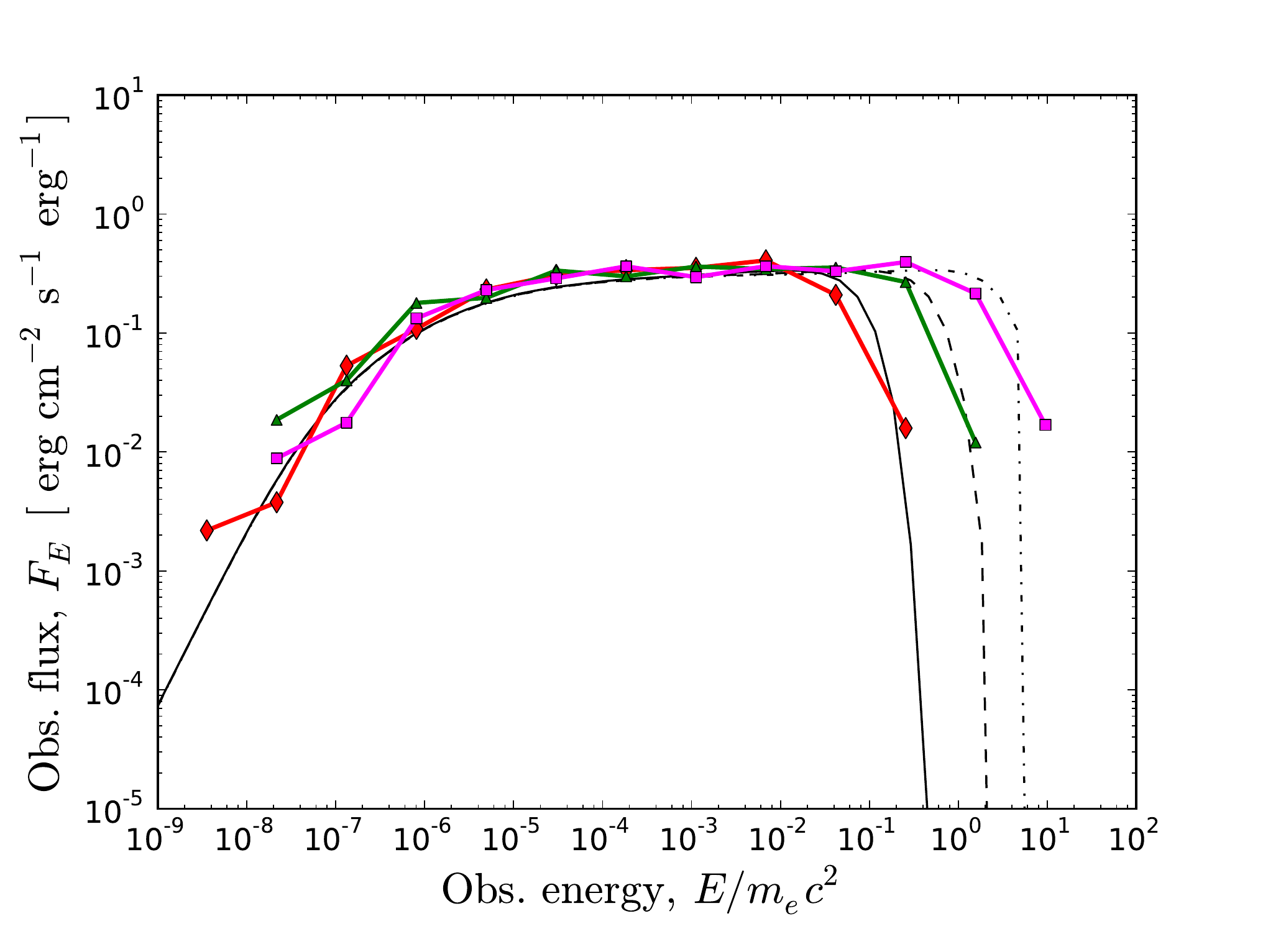}
\caption{Simulated (colored) and numerically integrated (black) spectra for narrow jets with different values of $\Gamma_{\rm 0}$ but constant values of $\theta_{\rm j} \Gamma_{\rm 0}$. In this plot, $p = 1$, $\theta_{\rm j} = 1/\Gamma_{\rm 0}$ and $\theta_{\rm v} = 0$ ($0.000 \leq \theta_{\rm v} \leq 0.006$ for red, $0.0000 \leq \theta_{\rm v} \leq 0.0032$ for green and $0.000 \leq \theta_{\rm v} \leq 0.0020$ for magenta). Three different values of $\Gamma_{\rm 0}$ are shown, $\Gamma_{\rm 0} = 100$ (red diamonds and solid black lines), $\Gamma_{\rm 0} = 200$ (green triangles and black dashed lines) and $\Gamma_{\rm 0} = 400$ (magenta squares and dash-dotted lines). After viewing angle binning, the red, green and magenta spectra contain $1934$, $558$ and $1114$ photons, respectively. The magenta spectrum requires small viewing angle bins due to the high $\Gamma_{\rm 0}$. Therefore, $1.3 \times 10^8$ photons were simulated for the magenta spectrum. For the other two spectra $1 \times 10^7$ photons were simulated. The spectral components from the inner and outer jet region are shifted to higher energies in accordance with Eqs. \ref{eq:E_peak} and \ref{eq:E_D}.}
\label{fig:p1v0j001_vary_g}
\end{figure}

The obtained spectra are presented in Figures \ref{fig:g100p1j001_vary_v} to \ref{fig:p1v0j001_vary_g}. In each figure we present both the simulated spectra and the numerical integration of Eq. \ref{eq:long spectrum}. We have considered a large set of the parameter space region: $\Gamma_{\rm 0} = 100, \, 200, \, 400$; $p = 1, \, 2, \, 4$; $\theta_{\rm j} \Gamma_{\rm 0} = 1, \, 3, \, 10$ and $\theta_{\rm v}/\theta_{\rm j} = 0, \, 0.5, \, 1, \, 1.5, \, 2$. In Figures \ref{fig:g100p1j001_vary_v} to \ref{fig:p1v0j001_vary_g} we have assumed non-dissipative fireball dynamics with $L = 10^{52} \, {\rm erg \, s^{-1}}$ and $r_{\rm 0} = 10^8 \, {\rm cm}$. A luminosity distance of $d_{\rm L} = 4.85 \times 10^{28} \, {\rm cm}$ ($z = 2$) was used for spectrum normalization. Each run contains $3 \times 10^7$ photons (unbinned) unless otherwise noted.

For a top-hat jet ($p \rightarrow \infty$), the most likely viewing angle is $\theta_{\rm v}/\theta_{\rm j} \approx 2/3$ since the jet is obscured for $\theta_{\rm v} > \theta_{\rm j}$. However, the outflows we consider here can be observed at $\theta_{\rm v} > \theta_{\rm j}$ due to the gradual decrease of the Lorentz factor. This increases the value of the most likely viewing angle. A value of $\theta_{\rm v} \approx \theta_{\rm j}$ may be close to the average value for jets with intermediate values of $p$ (the exact value of the most likely viewing angle depends both on the outflow as well as detector properties). Therefore we choose to show $\theta_{\rm v} / \theta_{\rm j} = 0, \, 1, \, 2$ in Figures \ref{fig:g100p1j001_vary_v} and \ref{fig:g100p4j010_vary_v}.

The most important result of this paper is that we obtain $\alpha \approx -1$ (where $\alpha$ is the low energy photon index in the Band function, $dN/dE \propto E^\alpha$ below $E_{\rm peak}$) for a large region of the considered parameter space. This is demonstrated in Figures \ref{fig:g100p1j001_vary_v} to \ref{fig:p1v0j001_vary_g}. We consider this a very important result as this is similar to the average low energy photon index observed in GRBs (e.g. \citealt{KanEtAl:2006, NavEtAl:2011, GolEtAl:2012}).

For a narrow jet ($\theta_{\rm j} = 1/\Gamma_{\rm 0}$, see Figure \ref{fig:g100p1j001_vary_v}), the peaks of the emission from the inner and outer jet regions ($E_{\rm peak}$ and $E_\mathcal{D}$ respectively, see discussion in \S \ref{sect:the observed spectrum}) coincides, and so the spectrum below $E_{\rm peak}$ is a pure power law for $\sim 6$ decades in energy. The resulting spectral shape is independent of viewing angle, however the peak energy decrease as the outflow is observed at large viewing angles.

In Figure \ref{fig:g100v1j001_vary_p} we present spectra for narrow jets with $p = 1, 2$ and $4$ observed at $\theta_{\rm v} = \theta_{\rm j}$. As can be seen, the low energy photon index is not very sensitive to the Lorentz factor gradient. The softest spectrum is obtained for $p = 1$, as shown in Figures \ref{fig:g100p1j001_vary_v} and \ref{fig:g100v1j001_vary_p}. For higher Lorentz factor gradients there is a slight increase in the photon index, consistent with the analytical expression in Eq. \ref{eq:spectrum solution 4}. For $1 < p < 4$, the photon index is $-1 \lesssim \alpha \lesssim -0.5$. Values of $p$ lower than unity has been considered. As $p$ decreases below unity the photon index below the peak energy increases. For $p = 0$ the outflow profile is spherically symmetric. Therefore the observed spectrum is that of a spherically symmetric wind ($\alpha \approx 0.4$).\footnote{For values of $p < 1$, the angle separating the outer jet region and the envelope ($\theta_{\rm e} \approx \theta_{\rm j} \Gamma_{\rm 0}^{1/p}$) becomes larger than unity (for $\theta_{\rm j} \geq \Gamma_{\rm 0}$) and the outflow consists only of the inner and outer jet regions.} For the simulated spectra, the exponential cut-off expected at energies above the peak energy becomes less sharp for increasing values of $p$. We discuss this further below.

The spectra from jets with large opening angles ($\theta_{\rm j} \approx 10/\Gamma_{\rm 0}$) observed at $\theta_{\rm v} \ll \theta_{\rm j}$ appear as those from spherically symmetric winds. However, for viewing angles $\theta_{\rm v} \approx \theta_{\rm j}$ the observed photon index below the peak energy is lower. In Figure \ref{fig:g100p4j010_vary_v} we present spectra from an outflow with the profile $\Gamma_{\rm 0} = 100$, $\theta_{\rm j} = 0.1$ and $p = 4$ observed at different viewing angles. For $\theta_{\rm v} = \theta_{\rm j}$, $\alpha \approx -1$ just as for narrow jets. Due to the large Lorentz factor gradient, $E_{\rm peak}$ decreases rapidly with increasing viewing angle. In Figure \ref{fig:g100p4j010_vary_v}, $E_{\rm peak}(\theta_{\rm v} = 2 \times \theta_{\rm j})/E_{\rm peak}(\theta_{\rm v} = 0) \approx 10^{-3}$. As discussed above, depending on the jet properties and detector characteristics the most likely viewing angle may be close to the jet opening angle.

In Figure \ref{fig:g100p2v0_vary_j} we present observed spectra from jets with different opening angles ($\theta_{\rm j} \Gamma_{\rm 0} = 1, 3$ and $10$) viewed head on. Both $\theta_{\rm j} \Gamma_{\rm 0} = 1, 3$ result in low energy slopes close to $\alpha = -1$ independent of viewing angle, while wider jets viewed at $\theta_{\rm v} = 0$ results in a spherically symmetric spectral shape a few decades below $E_{\rm peak}$.

Increasing the maximum Lorentz factor, $\Gamma_{\rm 0}$, shifts the spectral components from the inner and outer jet up in energy while keeping their spectral shapes intact (Figure \ref{fig:p1v0j001_vary_g}, also see Eqs. \ref{eq:E_peak}, \ref{eq:E_j}, \ref{eq:E_D} and \ref{eq:E_e}). The envelope component is unaffected (as expected, see Eq. \ref{eq:E_env}). As long as $\Gamma_{\rm 0} \theta_{\rm j}$ is constant, the spectral shapes of the inner and outer jet components are not affected by varying $\Gamma_{\rm 0}$.

We reach the conclusion that the low energy spectral index is close to $\alpha \approx -1$ for narrow jets, with $\theta_{\rm j} \leq few/\Gamma_{\rm 0}$ and moderate Lorentz factor gradients for all viewing angles. Similar photon indices are obtained from wider jets observed at $\theta_{\rm v} \approx \theta_{\rm j}$.

\begin{figure}
\includegraphics[width=\linewidth]{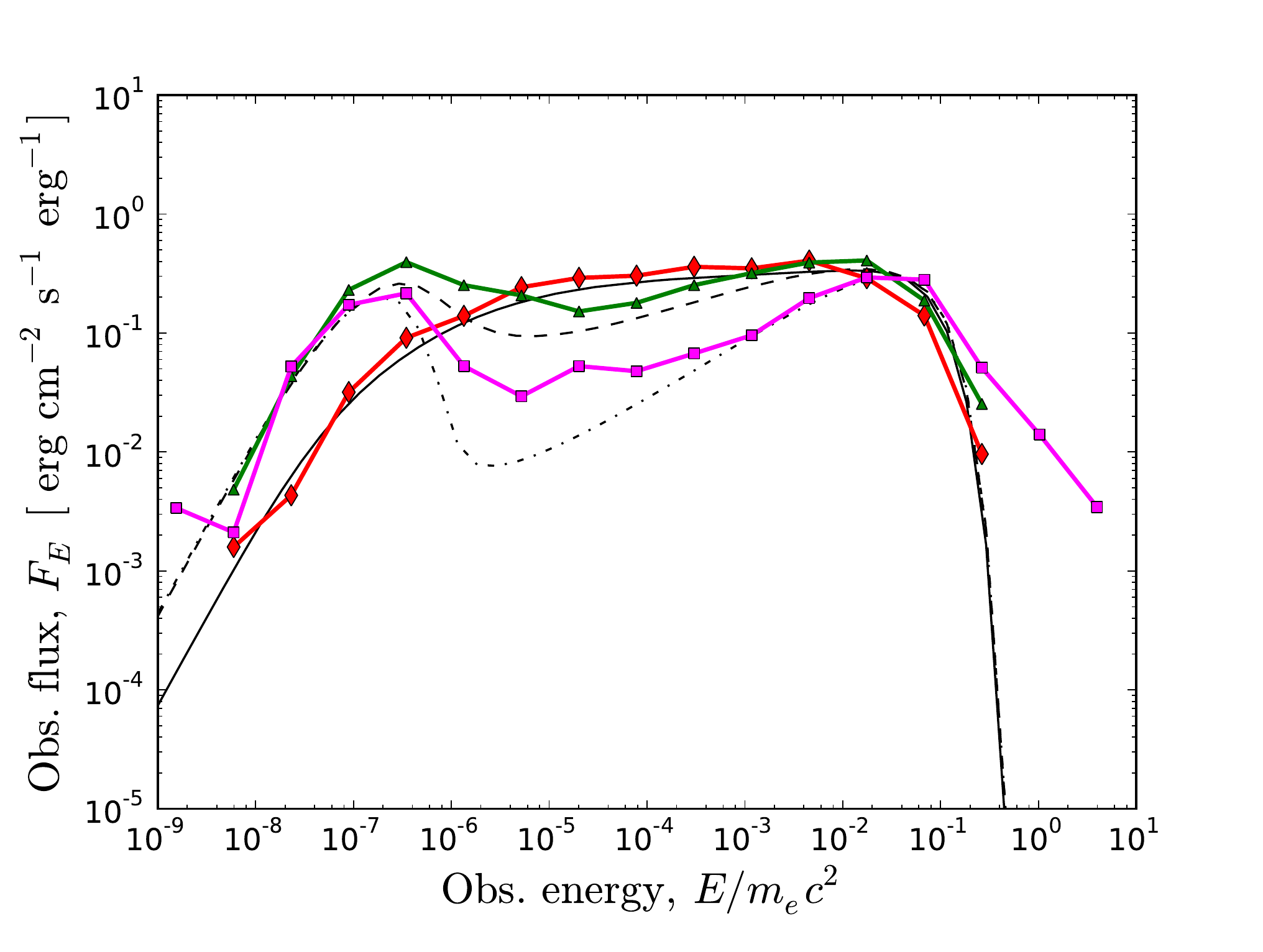}
\caption{Same as Figure \ref{fig:g100v1j001_vary_p}, but $\theta_{\rm v} = 0$ ($0.000 \leq \theta_{\rm v} \leq 0.006$). After viewing angle binning, the red, green and magenta spectra contain $1934$, $2166$ and $1178$ photons, respectively. A power law above the peak energy is seen in the magenta spectrum. The power law is further discussed in the text.}
\label{fig:g100v0j001_vary_p}
\end{figure}

\subsection{Asymmetric photon diffusion and Comptonization}

Preferential photon diffusion towards the jet center is expected to occur in outflows where the bulk Lorentz factor decreases from the jet axis. This can be understood in terms of average photon scattering angles. The average photon scattering angle with respect to the radial direction is $\sim 1/\Gamma$ in the lab frame. Therefore a photon is more likely to scatter from a region of low Lorentz factor to a region of high Lorentz factor than the other way around. The importance of photon diffusion is dependent on the Lorentz factor gradient, and so the spectra from jet profiles with large values of $p$ are expected to differ from the numerically integrated spectra where we assume $d\dot{N}_\gamma/d\Omega$ is $r$-independent. This is indeed observed in the simulations. The effects of angular photon diffusion are, however, sub-dominant as compared to the observed temperature variations across the jet when considering the formation of the observed spectrum from jets with small Lorentz factor gradients (approximately $p \lesssim 4$).

Comptonization of photons propagating to regions of higher Lorentz factor may cause photons to fall out of thermal equillibrium with the local electrons. This has been investigated by comparing the simulated lab frame photon energy to the lab frame temperature of the electrons at the last scattering position for all photons in a given spectrum. The observed photon energies are well described by the local observed temperature for most of the considered parameter space. However, a deviation is observed in Figures \ref{fig:g100v1j001_vary_p} and \ref{fig:g100v0j001_vary_p}. The expected high energy cut-off is significantly smoother for the spectrum with $p=4$ observed at $\theta_{\rm v} = \theta_{\rm j}$. The same jet observed at $\theta_{\rm v} = 0$ forms a high energy power law with photon index $\beta \approx -2$. The photons forming the power law start out with thermal energies in the outer jet region where the Lorentz factor gradient is large. Repeated scatterings towards the jet center increases the photon energy until the photon reaches the inner jet region where it finally escapes. The final photon temperature depends on the initial thermal photon energy, the number of scatterings before escape and the average energy gained in each scattering. The last two terms are dependent on the Lorentz factor gradient. The Comptonization of photons propagating in media with a Lorentz factor gradient will be further explored in future works. In general, the analytical expressions for characteristic energies and spectral shapes presented in \S \ref{sect:the observed spectrum} agree well with the Monte Carlo simulations except for large Lorentz factor gradients.

%%%%%%%%%%%%%%%%%%%%%%%%%%%%%%%%%%%%%%%%%%%%%%%%
%%%%%%%%%%%%%%%%%%%%%%%%%%%%%%%%%%%%%%%%%%%%%%%%
%%%%%%%%%%%%%%%%%%%%%%%%%%%%%%%%%%%%%%%%%%%%%%%%
%%%%%%%%%%%%%%%%%%%%%%%%%%%%%%%%%%%%%%%%%%%%%%%%
%%%%%%%%%%%%%%%%%%%%%%%%%%%%%%%%%%%%%%%%%%%%%%%%
%%%%%%%%%%%%%%%%%%%%%%%%%%%%%%%%%%%%%%%%%%%%%%%%
%%%%%%%%%%%%%%%%%%%%%%%%%%%%%%%%%%%%%%%%%%%%%%%%
%%%%%%%%%%%%%%%%%%%%%%%%%%%%%%%%%%%%%%%%%%%%%%%%
%%%%%%%%%%%%%%%%%%%%%%%%%%%%%%%%%%%%%%%%%%%%%%%%
%%%%%%%%%%%%%%%%%%%%%%%%%%%%%%%%%%%%%%%%%%%%%%%%

\section{Discussion}
\label{sect:discussion}

\subsection{Sensitivity to the chosen dynamics and Lorentz factor profile}

In this work we consider fireball dynamics for the outflow properties. However, other dynamics are possible: kinetic energy dissipation may be significant during the acceleration phase, or the outflow may be dominated by magnetic fields. The overall shape of the spectral components from the inner jet and evelope regions are not expected to be sensitive to the exact dynamics, however the value of $E_{\rm peak}$ is dependent on the radial scalings of the outflow properties. The exact value of the power law slope in the outer jet component is expected to be modified for other dynamics, as well as $E_\mathcal{D}$ (Eq. \ref{eq:E_D}, the energy of the spectral bump below the peak energy in Figure \ref{fig:general spectrum}). However, the ratio $E_{\rm j}/E_{\rm peak}$ which gives an indication of the energy where the spectrum starts to deviate from the sperically symmetric spectrum is not sensitive to the dynamics.

The shape of the angular profile is motivated by hydrodynamical simulations. One may consider other shapes (e.g. Gaussian) for the angular parameters. The exact values of the characteristic energies and spectral shapes are naturally dependent on the chosen profile, however the general spectral features are not sensitive as long as the considered profile has similar characteristics.

\subsection{Relative photon time delays}
\label{subsect:time delays}

In this work we focus on the spectrum from steady jets. In much the same way as photons from a certain angle has a characteristic energy, they also have a characteristic time delay as compared to a photon from the center of the jet. One can therefore associate a given photon energy in the spectrum with a certain time delay.

Assuming the jet is viewed at zero viewing angle, the characteristic lab frame time delay as a function of angle is

\begin{eqnarray}
\Delta t(\theta) & \approx & \frac{R_{\rm ph}(\theta)}{c}\left[1 - \beta(\theta)\cos\theta\right] - \frac{R_{\rm ph}(0)}{c}\left[1 - \beta(0)\right] \nonumber \\
                 & \approx & \frac{R_{\rm ph}(\theta)}{c}\left[1 - \beta(\theta)\cos\theta\right]
\end{eqnarray}

\noindent where $\theta > 1/\Gamma_{\rm 0}$ is assumed in the last step. The time delay for a given energy is generally larger for an outflow with a Lorentz factor that decreases from the jet axis, as compared to the sperically symmetric case. It increases with decreasing $\theta_{\rm j}$ and increasing $p$, but decreases rapidly with $\Gamma_{\rm 0}$. As an upper limit (i.e. the largest time delay for a given energy), we consider a jet profile with $\theta_{\rm j} = 1/\Gamma_{\rm 0}$. The time delay of photons with lower energy than the observed peak energy is calculated as follows. { Consider photons that make their last scattering in the outer jet region (i.e. at angles $\theta > 1/\Gamma_{\rm 0}$, since $\theta_{\rm j} = 1/\Gamma_{\rm 0}$)}. For $\theta_{\rm j} = 1/\Gamma_{\rm 0}$, $\theta_\mathcal{D} = \theta_{\rm j}$ and Eqs. \ref{eq:gamma pl}, \ref{eq:doppler pl} and \ref{eq:r_ray pl} are good approximations for $\Gamma(\theta)$, $\mathcal{D}(\theta)$ and $R_{\rm ph}(\theta)$ respectively. The characteristic photon energy at angle $\theta$ is $E(\theta)/E_{\rm peak} \approx \{[\mathcal{D}(\theta)/\Gamma(\theta)] [r_{\rm s}(\theta)/R_{\rm ph}(\theta)]^{2/3}\}/\{2 [r_{\rm s}(0)/R_{\rm ph}(0)]^{2/3}\}$. Since $\Delta t \approx R_{\rm ph} (1 - \beta\cos\theta)/c = R_{\rm ph}/\mathcal{D} \Gamma c$, we write

\begin{eqnarray}
\Delta t^{\rm ob} & \approx & 1.3 \left(\frac{1+z}{3}\right) \left(\frac{L}{10^{52}}\right) \left(\frac{\Gamma_{\rm 0}}{300}\right)^{-5} \times \nonumber \\
& & \left(\frac{E/E_{\rm peak}}{10^{-2}}\right)^{-15/8} \left(\frac{3p+1}{13}\right)^{1/4} \, {\rm s}.
\label{eq:time delay}
\end{eqnarray}

\noindent { which is valid for $E \ll E_{\rm peak}$, $\theta_{\rm v} \ll \theta_{\rm j}$ and $\theta_{\rm j} = 1/\Gamma_{\rm 0}$}. Time-resolved analysis of spectra detected down to two orders of magnitude below the peak energy may resolve the superposition of blackbodies in the lowest observed energies if the time bins are smaller than $\Delta t^{\rm ob}$. The result in Eq. \ref{eq:time delay} shows that for a narrow jet with otherwise typical GRB parameters, the expected time delay is approximately $1 \, {\rm s}$ for photons with energy $E \approx 10^{-2} \, E_{\rm peak}$. { For wider jets ($\theta_{\rm j} > 1/\Gamma_{\rm 0}$) the time delay is shorter.

This result may be compared to the case of an angle independent emission radius (such as the case for the afterglow, e.g. \citet{KumPan:2000}). Assuming the emission radius to be equal to the photospheric radius at the LOS, photons with energy $E = 10^{-2} \, E_{\rm peak}$ (originating at $\theta > 1/\Gamma$) arrives with a time delay of $\Delta t^{\rm ob} \approx 4 \times 10^{-3} (L/10^{52}) (\Gamma/300)^{-5} ([E/E_{\rm peak}] / 10^{-2})^{-1} \, {\rm s}$. This time delay is three orders of magnitude shorter than the result obtained in Eq. \ref{eq:time delay}. This example highlights the importance of recognizing the angular dependence of the photospheric radius.}

The spectral component associated with the envelope has a very long associated time delay, due to the large photospheric radius (see Figure \ref{fig:approximations}). We therefore do not expect the envelope component to be observed in the prompt emission emitted from a transient source.

\subsection{Band-like GRB spectra}

Observed GRB spectra do not, in general, appear thermal. The spectra are usually well fitted by the Band function, although more complex spectra has been observed (e.g. \citealt{RydEtAl:2010, AckEtAl:2011, GuiEtAl:2011}). The average values of the low and high energy photon indices are $\alpha \approx -1$ and $\beta \approx -2.5$ respectively. This is in contrast to a blackbody spectrum in which $\alpha = 1$ and the spectrum cuts off exponentially above the peak energy. To reconcile the photospheric model with observations, energy dissipation close to the photosphere has been considered as a way to modify the local comoving photon spectrum \citep{Tho:1994, SprDaiDre:2001, ReeMes:2005, PeeMesRee:2005, GiaSpr:2005, PeeMesRee:2006, LazMorBeg:2009, Bel:2010, RydEtAl:2011, ZhaYan:2011, VurEtAl:2011}. In these scenarios the observed spectrum results from a combination of radiative processes. Comptonization of the thermal photon spectrum is the source of the high energy power law, and synchrotron photons contribute to the low energy spectrum (e.g. \citealt{VurEtAl:2011}). In this work we show that three dimensional effects arising from the jet shape may add a significant contribution to understanding the soft low energy spectrum. Narrow jets, or wider jets observed at $\theta_{\rm v} \approx \theta_{\rm j}$ result in a low energy photon index $\alpha \approx -1$ under natural assumptions. Harder values of the low energy photon index are obtained for wide jets viewed at $\theta_{\rm v} \ll \theta_{\rm j}$. { The details of the observed spectrum predicted here are expected to be modified if significant energy dissipation takes place below the photosphere, since in this scenario the outflow dynamics is changed. Moreover, as discussed above, in such a scenario the soft tail of the spectrum may be modified by non-thermal emission.}

In some GRBs observed by the {\it Fermi Gamma-Ray Space Telescope} (GRB 100724B, \citealt{GuiEtAl:2011} and GRB 110721A, \citealt{AxeEtAl:2012}), a spectral ``bump'' is significantly detected at energies below the main peak. This bump is commonly interpreted as photospheric emission (e.g. \citealt{Ryd:2005}). In such a scenario, the Band component originates from non-thermal emission processes in the optically thin part of the outflow outside the photosphere (e.g. \citealt{ZhaEtAl:2012, VerZhaMes:2012}). Within this interpretation the analysis performed in this paper is applicable to the bump instead of the Band component and may explain bumps wider than blackbody.

%%%%%%%%%%%%%%%%%%%%%%%%%%%%%%%%%%%%%%%%%%%%%%%%
%%%%%%%%%%%%%%%%%%%%%%%%%%%%%%%%%%%%%%%%%%%%%%%%
%%%%%%%%%%%%%%%%%%%%%%%%%%%%%%%%%%%%%%%%%%%%%%%%
%%%%%%%%%%%%%%%%%%%%%%%%%%%%%%%%%%%%%%%%%%%%%%%%
%%%%%%%%%%%%%%%%%%%%%%%%%%%%%%%%%%%%%%%%%%%%%%%%
%%%%%%%%%%%%%%%%%%%%%%%%%%%%%%%%%%%%%%%%%%%%%%%%
%%%%%%%%%%%%%%%%%%%%%%%%%%%%%%%%%%%%%%%%%%%%%%%%
%%%%%%%%%%%%%%%%%%%%%%%%%%%%%%%%%%%%%%%%%%%%%%%%
%%%%%%%%%%%%%%%%%%%%%%%%%%%%%%%%%%%%%%%%%%%%%%%%
%%%%%%%%%%%%%%%%%%%%%%%%%%%%%%%%%%%%%%%%%%%%%%%%

\section{Summary and Conclusions}
\label{sect:summary and conclusions}

In this work we develop the theory of photospheric emission from relativistic jets with angle dependent outflow properties. We consider a three parameter angular Lorentz factor profile (Eq. \ref{eq:gamma profile}) where the Lorentz factor is approximately constant, $\Gamma = \Gamma_{\rm 0}$ within a characteristic jet opening angle $\theta_{\rm j}$ and then decreasing approximately as $\Gamma \propto \theta^{-p}$ towards the outer jet edge. The shape of the profile is motivated by the results of hydrodynamical simulations of jet propagation through the progenitor star (e.g. \citealt{AloEtAl:2000, ZhaWooMac:2003, MorLazBeg:2007, MizNagAoi:2011}). In \S \ref{sect:model} the expression for the observed spectrum is obtained analytically by integrating the local emissivity over all radii and angles (Eq. \ref{eq:long spectrum}). We derive approximate analytical expressions for the important spectral features in \S \ref{sect:the observed spectrum}. Comparing jetted outflows to spherical outflows, we show that softening of the spectrum below the peak energy is expected for an on-axis observer. This is a consequence of weaker radial beaming of photons at high latitudes due to the decreasing Lorentz factor. A Monte Carlo simulation was developed to investigate the importance of full photon propagation below the photosphere (\S \ref{sect:numerical simulations}). In \S \ref{sect:results} we present spectra obtained by numerical integration of Eq. \ref{eq:long spectrum} as well as the Monte Carlo simulations for different profile parameters and viewing angles (Figures \ref{fig:g100p1j001_vary_v} to \ref{fig:g100v0j001_vary_p}).

The most important result of this paper is that { the photospheric spectrum below the thermal peak may be significantly softer than blackbody, as a consequence of geometrical broadening. In particular,} we obtain a photon index $\alpha \approx -1$ below the peak energy for a large region of the considered parameter space. For narrow jets ($\theta_{\rm j} \lesssim few/\Gamma_{\rm 0}$) with Lorentz factor gradients $1 \leq p \leq 4$ observed at any viewing angle, we find $-1 \gtrsim \alpha \gtrsim -0.5$. For jets with $\theta_{\rm j} \approx 1/\Gamma_{\rm 0}$ and $p \geq 1$, $\alpha \approx -(1/4)(1 + 3/p)$ (see Eq. \ref{eq:spectrum solution 4}). Observing wider jets ($\theta_{\rm j} \gtrsim 5/\Gamma_{\rm 0}$) at $\theta_{\rm v} \approx \theta_{\rm j}$ results in similar soft spectra with $\alpha \approx -1$. { However, spectra from wide jets observed at small angles ($\theta_{\rm v} \ll \theta_{\rm j}$) appears similar to the spectrum from a spherical wind (i.e. close to blackbody but with $\alpha \approx 0.4$ at $E = 10^{-2} \, E_{\rm peak}$) for several decades below the peak energy. This may explain the hard low energy photon indices observed in some GRBs ($\alpha \approx 0$ \citep{GolEtAl:2012}, see further discussion below).} However, observing the outflow at viewing angles $\theta_{\rm v} \approx 0$ is less likely than $\theta_{\rm v} \approx \theta_{\rm j}$. Additionally, increasing the viewing angle causes the observed peak energy to decrease. The decrease is slower for low Lorentz factor gradients. Therefore, such outflows may still be observed at $\theta_{\rm v} \approx few \times \theta_{\rm j}$.

Photon diffusion primarily towards regions of higher Lorentz factor is observed in the simulations. For outflows with large Lorentz factor gradient ($p \gtrsim 4$) this bulk propagation of photons has to be taken into account when computing the observed spectrum. Comptonization of the photons that make repeated scatterings towards regions of larger Lorentz factor can produce photons with energies significantly higher than the local temperature. Evidence for this can be seen in Figure \ref{fig:g100v0j001_vary_p} ($p = 4$), where an approximate high energy power law is formed above the peak energy. Photon propagation in plasma with a steep Lorentz factor gradient will be further explored in future works.

The observed photospheric spectrum from a spherical outflow has previously been considered in the literature (e.g. \citealt{Bel:2010, PeeRyd:2011}). In the limiting case of outflows with $\theta_{\rm j} \gtrsim 5/\Gamma_{\rm 0}$ observed at $\theta_{\rm v} \ll \theta_{\rm j}$, we obtain similar results. For larger viewing angles or smaller jet opening angles, geometrical broadening of the spectrum has to be considered.

Although we consider a static source, as shown in \S \ref{subsect:time delays} there is a time delay associated with high-latitude photons. The time delay increases with decreasing observed photon energy. The time delay at a specific energy is longer for more narrowly collimated jets. For a narrow jet ($\theta_{\rm j} = 1/\Gamma_{\rm 0}$) with typical outflow parameters characterizing GRBs, the time delay is $\approx 1 \, {\rm s}$ at $E \approx 10^{-2} E_{\rm peak}$. Spectral analysis of prompt GRB emission using smaller time bins than this time delay may reveal harder spectra within the energy range $10^{-2} E_{\rm peak} \lesssim E \lesssim E_{\rm peak}$ than what is predicted by the static model.

For this work we have considered outflows with angle independent luminosity, and so the shape of the Lorentz factor profile is fully determined by the angle dependent baryon loading. As shown above, the spectral slope in the range $10^{-2} E_{\rm peak} \lesssim E \lesssim E_{\rm peak}$ is formed by photons making their last scattering at $\theta \lesssim 5/\Gamma_{\rm 0}$. Therefore, this assumption is a good approximation for jets where $dL/d\Omega \approx const$ for $\theta \lesssim 5/\Gamma_{\rm 0}$. This requirement is fulfilled for model JA in \citet{ZhaWooMac:2003} close to the largest radius of the simulation ($r = 2.1 \times 10^{10} \, {\rm cm}$). In our notation, $\Gamma_{\rm 0} \approx 140$, $\theta_{\rm j} \approx 0.017 \approx 2.5/\Gamma_{\rm 0}$ and $p \approx 6.5$ for model JA as estimated from Figures 8 and 9 in \citet{ZhaWooMac:2003}.

Furthermore, we consider non-dissipative fireball dynamics. The dynamics of the outflow are dependent on the dominant form of energy carried by the jet as well as energy dissipation. In particular, the radial scalings of the comoving temperature and Lorentz factor in magnetically dominated outflows are expected to differ significantly from the scalings of thermal fireballs. \citet{Gia:2012} and \citet{Bel:2012} considered the scaling of $E_{\rm peak}$ in dissipative outflows. Heating keeps the comoving temperature approximately constant in the range $R_{\rm ph}/30 \lesssim r \lesssim R_{\rm ph}$, in contrast to the non-dissipative outflows considered here. However, the framework presented in \S \ref{sect:model} is general and may be applied to any relativistic, optically thick outflow.

Since the causally connected parts of the outflow are separated by angles $\approx 1/\Gamma$, one may consider outflows with Lorentz factor variations at angular scales of $\Delta\theta \approx few/\Gamma$. Geometrical broadening of the observed spectrum is expected as a consequence of the beaming of photons being a function of angle from the jet axis, in much the same way as for the jets considered in this work. The spectral slope below the peak energy is expected to depend on both the typical angular scale as well as the amplitude of the Lorentz factor variations.

Spectral broadening by energy dissipation in regions of moderate optical depth may be combined with geometrical broadening. As the observed spectrum below the peak energy is a superposition of comoving spectra, it is not sensitive to the exact shape of the comoving spectrum. Comptonization of the comoving spectrum by electrons which are heated by energy dissipation (e.g. \citealt{Bel:2010, LazBeg:2010}) can shape the spectrum above the peak energy, while geometrical broadening forms the spectrum below the peak energy.

The observed low energy photon index varies between bursts, forming an approximately Gaussian distribution centered at $\alpha \approx -1$ with a full width at half maximum of $\approx 1$ (e.g. \citealt{GolEtAl:2012}). Within the framework presented in this paper, the distribution could naturally be interpreted as a result of viewing angle variations. { In particular, observing a wide jet at zero viewing angle results in $\alpha \approx 0.4$.} The exact $\alpha$-distribution predicted by our model is hard to obtain, since it depends on detector characteristics. However, a clear prediction of the model is that of softening of the low energy photon index with increasing viewing angle for jets with $\theta_{\rm j} \gtrsim few/\Gamma_{\rm 0}$.

GRB 090902B is a burst of special interest, due to its unusual spectral shape \citep{AbdEtAl:2009}. The prompt spectrum consists of a sharply peaked Band component along with a wide power law. The hard low energy photon index and the narrow width of the Band component poses an extreme challenge for optically thin emission models. Therefore, a photospheric origin of the Band component in this burst seems inevitable \citep{RydEtAl:2010, RydEtAl:2011, ZhaEtAl:2011, PeeEtAl:2012}. Since the Band component appears close to blackbody, processes that are expected to modify the photospheric spectrum can be constrained for GRB 090902B. The rate of energy dissipation in regions of moderate optical depth must be relatively low as the observed Band component is not severely distortened from the Planck spectrum. Within the framework considered in this paper, further constraints can be set. The requirement for geometrical broadening to not significantly distort the observed spectrum in the energy range $10^{-2} E_{\rm peak} \lesssim E \lesssim E_{\rm peak}$ is $\theta_{\rm v} \ll \theta_{\rm j}$ and $\theta_{\rm j} \gtrsim 5/\Gamma_{\rm 0}$. Depending on typical GRB outflow characteristics, the probability for such parameter combinations may be low. This would help explaining the rarity of similar events.

Multiple spectral components have been clearly identified in several GRBs after the launch of the {\it Fermi} telescope \citep{AbdEtAl:2009, AckEtAl:2010, AckEtAl:2011}. This was expected from a theoretical point of view, simply because of the wide energy range of the observations. Several models predict non-thermal spectral components as a result of kinetic or magnetic energy dissipation above the photosphere (e.g. \citealt{MesEtAl:2002, ZhaYan:2011}). If dissipation occurs at $r \gg R_{\rm ph}$ the thermal and non-thermal components may be fitted separately (such as in 090902B \citep{PeeEtAl:2012}). For dissipation at $r \gtrsim R_{\rm ph}$ the separation of thermal and non-thermal components may be less clear because of the coupling of the thermal photon field to the accelerated non-thermal electrons. In this work we consider the thermal component in isolation in order to demonstrate the geometrical broadening of the spectrum. However, a complete theory for the prompt GRB emission must explain all observed spectral components.

An outstanding problem for the photospheric interpretation of prompt GRB emission is the non-thermal spectra commonly observed. In particular, the average low energy photon index is not compatible with the Rayleigh-Jeans index ($\alpha \approx -1$ as compared to $\alpha = 1$). In this work we show that either a narrow jet ($\theta_{\rm j} \lesssim few / \Gamma_{\rm 0}$) with a moderate Lorentz factor gradient ($1 \leq p \leq 4$) observed at any viewing angle or a wide jet observed at $\theta_{\rm v} \approx \theta_{\rm j}$ can naturally produce $-1 \lesssim \alpha \lesssim -0.5$ through geometrical broadening, without a need for additional emission processes such as synchrotron emission to supply photons below the thermal peak.

\section*{Acknowledgments}
{ We would like to thank Peter M{\'e}sz{\'a}ros, Martin Rees and Bing Zhang for useful discussions.} The Swedish National Space Board is acknowledged for financial support. AP wishes to acknowledge support from Fermi GI program \#41162.

%%%%%%%%%%%%%%%%%%%%%%%%%%%%%%%%%%%%%%%%%%%%%%%%
%%%%%%%%%%%%%%%%%%%%%%%%%%%%%%%%%%%%%%%%%%%%%%%%
%%%%%%%%%%%%%%%%%%%%%%%%%%%%%%%%%%%%%%%%%%%%%%%%
%%%%%%%%%%%%%%%%%%%%%%%%%%%%%%%%%%%%%%%%%%%%%%%%
%%%%%%%%%%%%%%%%%%%%%%%%%%%%%%%%%%%%%%%%%%%%%%%%
%%%%%%%%%%%%%%%%%%%%%%%%%%%%%%%%%%%%%%%%%%%%%%%%
%%%%%%%%%%%%%%%%%%%%%%%%%%%%%%%%%%%%%%%%%%%%%%%%
%%%%%%%%%%%%%%%%%%%%%%%%%%%%%%%%%%%%%%%%%%%%%%%%
%%%%%%%%%%%%%%%%%%%%%%%%%%%%%%%%%%%%%%%%%%%%%%%%
%%%%%%%%%%%%%%%%%%%%%%%%%%%%%%%%%%%%%%%%%%%%%%%%

\bibliographystyle{mn2e}
\bibliography{references}

\label{lastpage}

\end{document}